\documentclass[12pt,amsmath,amssymb,aps]{revtex4-1}

\usepackage{graphicx} 
\usepackage{color} 
\usepackage{bm} 

\newcommand{\blue}{\textcolor{blue}}
\newcommand{\red}{\textcolor{red}}
\newcommand{\green}{\textcolor{green}}

\begin{document}

\title{Nanodrop impact on solid surfaces}

\author{Joel Koplik}
\email{koplik@sci.ccny.cuny.edu}
\author{Rui Zhang}
\email{ruizhang@ccny.cuny.edu}
\affiliation{Benjamin Levich Institute and Department of Physics \\
City College of the City University of New York, New York, NY 10031}

\date{\today}

\begin{abstract}
The impact of nanometer sized drops on solid surfaces is studied
using molecular dynamics simulations.  Equilibrated floating drops
consisting of short chains of Lennard-Jones liquids with
adjustable volatility are directed normally onto an atomistic solid
surface where they are observed to bounce, stick, splash or disintegrate,
depending on the initial velocity and the nature of the materials involved.
Drops impacting at low velocity bounce from non-wetting surfaces but stick
and subsequently spread slowly on wetting surfaces.
Higher velocity impacts produce an prompt splash
followed by disintegration of the drop, while at still higher velocity drops
disintegrate immediately.  The disintegration can be
understood as either a loss of coherence of the liquid or as the result of a
local temperature exceeding the liquid-vapor coexistence value.  In contrast to
macroscopic drops, the presence of vapor outside the drop does not effect the
behavior in any significant way. Nonetheless, the transition between the 
splashing and 
bouncing/sticking regimes occur at Reynolds and Weber numbers similar to 
those found for larger drops.
\end{abstract}

\maketitle

\section{Introduction}
\label{sec:intro}
The splashing of liquid drops on a solid surface has long been a subject of 
both visual appreciation \cite{worth} and quantitative fluid mechanical
studies \cite{rein,yariv}.  Recent experiments \cite{xu} have reinvigorated
this subject, by showing that removal of the vapor surrounding a drop
suppresses splashing, an effect commonly associated with the presence of a
lubrication layer of gas beneath an impacting drop, and a number of papers
discuss splash modeling taking account of the gas. Recent examples, from
which earlier literature may be traced, include \cite{mani,hicks,duchemin}.  
Detailed flow measurements of splashing laboratory drops are challenging, 
due to the speed of the process and the small length scales involved, which
motivates us to perform molecular dynamics simulations. Here, control of the 
materials and initial conditions is straightforward and configurational 
information and flow fields at fine spatial and temporal resolution are 
relatively easy to obtain.  A particular advantage of this method is that
the initial presence and distribution of vapor outside the drop may be
altered at will.  The price of high resolution and control is that only
relatively small systems may be studied in this way, and we have considered
drops roughly 12 nm in radius.  Although many aspects of macroscopic fluid
flows have been found to persist down to this scale, differences exist (see
below), and we find that nanodrop impacts only partially agree with recent
laboratory observations.  In particular, the presence of vapor does not
appear to be necessary for nanodrops to splash.  
   
\section{Simulation method}
\label{sec:meth}
The simulations employ standard classical molecular dynamics 
techniques \cite{at,fs}.  The drops are composed of small liquid molecules
containing either two or four Lennard-Jones (LJ) atoms, bound into flexible 
linear chains by a FENE interaction \cite{fene}. The drops contain 157,126 
atoms placed at the center of a box of dimensions $(X,Y,Z) = 
(300,100,300)\sigma$ and are initially equilibrated while floating freely  
using a Nos\'e-Hoover thermostat at temperature
T=0.8 $\epsilon/k_B$.  Here, $\epsilon$ and $\sigma$ are the length and
energy parameters in the LJ potential, and $k_B$ is Boltzmann's
constant. The characteristic time unit in the simulations is 
$\tau=\sigma(m/\epsilon)^{1/2}$ where $m$ is the atomic mass.
Representative  physical values of these parameters are $\sigma \sim 0.34$nm 
and $\tau \sim 2$ ps. In the simulations, Newton's equations are integrated
using a fifth-order predictor-corrector method with a time step of
0.0005$\tau$.

\begin{figure}[t]
  \includegraphics[width=0.19\linewidth]{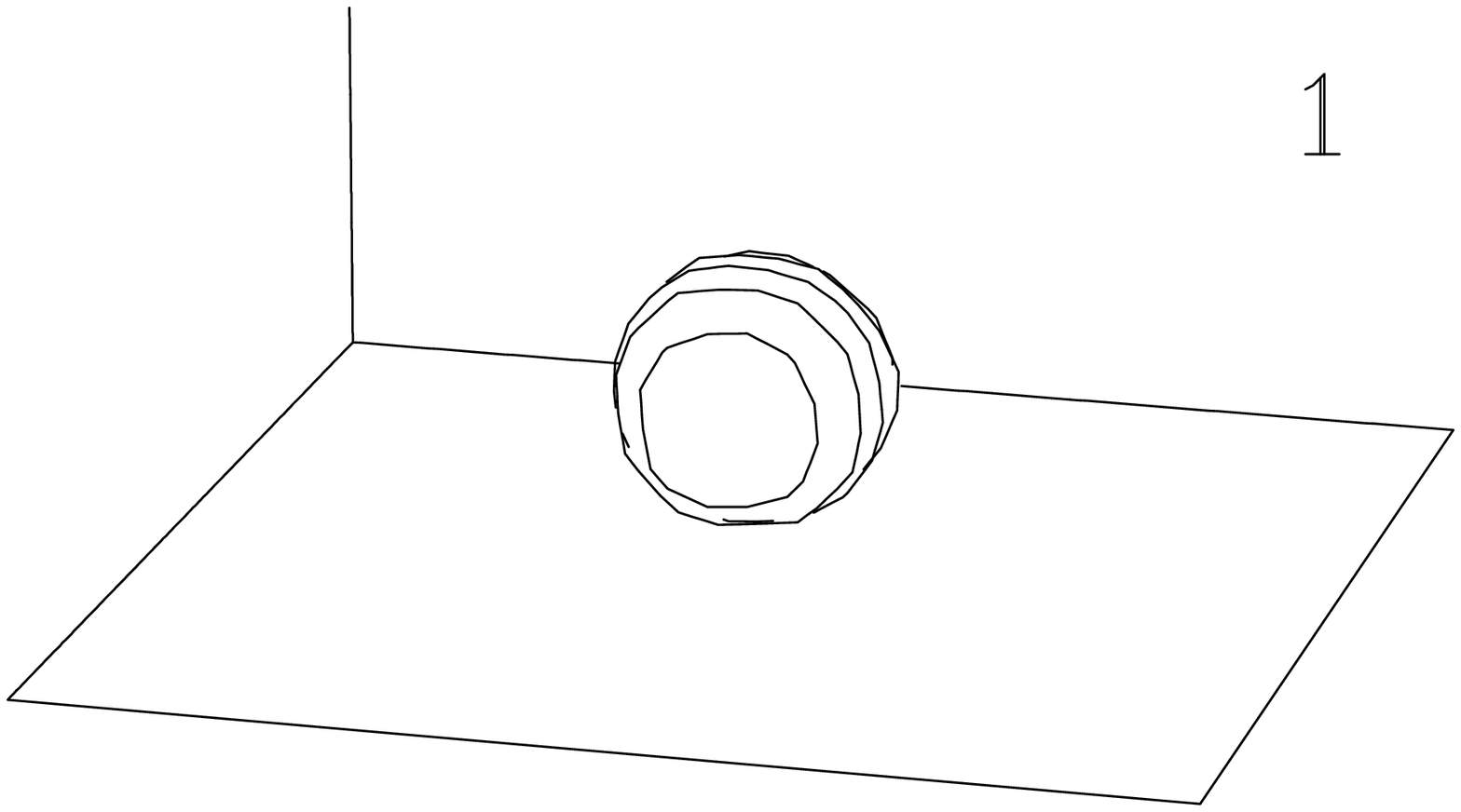}
  \includegraphics[width=0.19\linewidth]{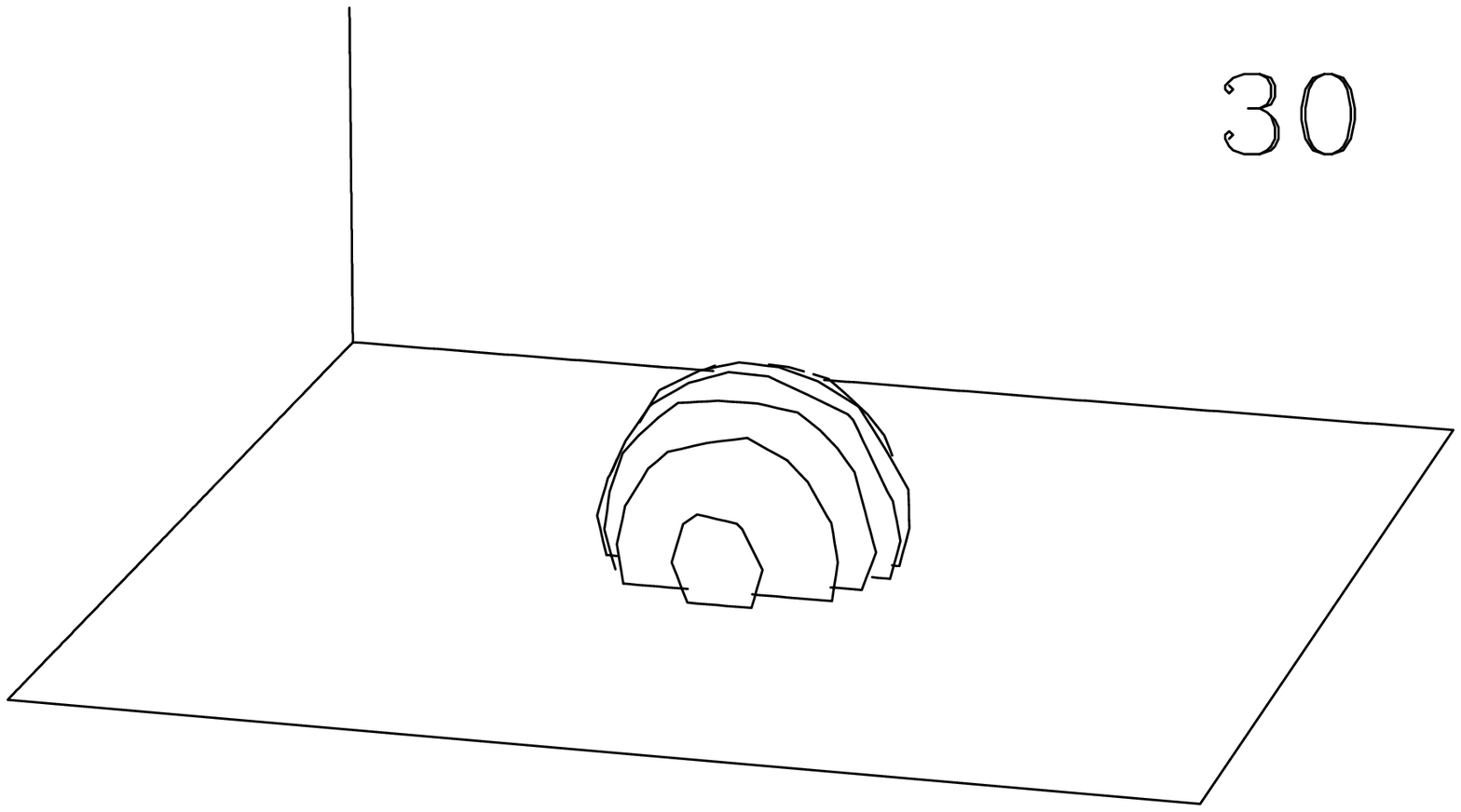}
  \includegraphics[width=0.19\linewidth]{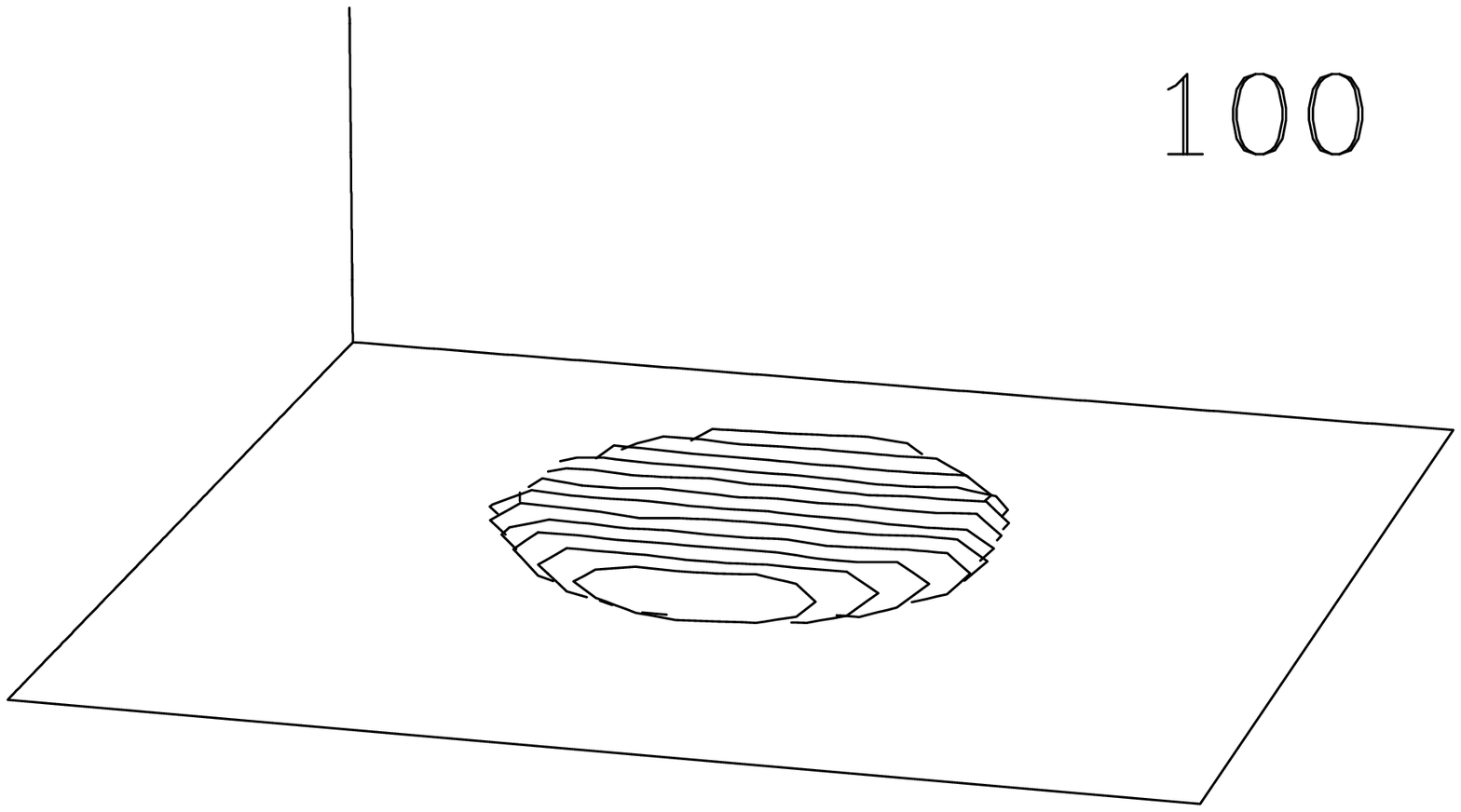}
  \includegraphics[width=0.19\linewidth]{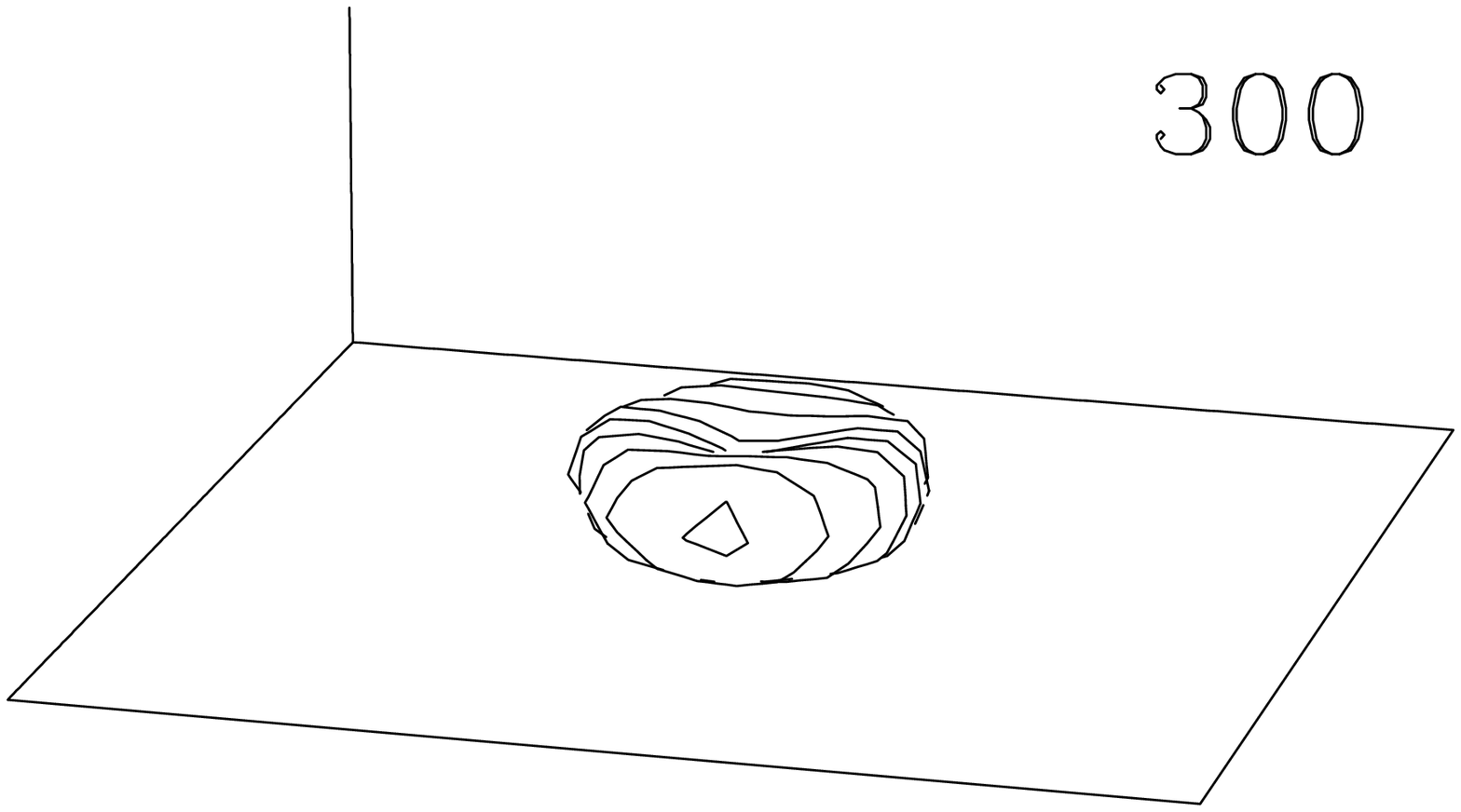}
  \includegraphics[width=0.19\linewidth]{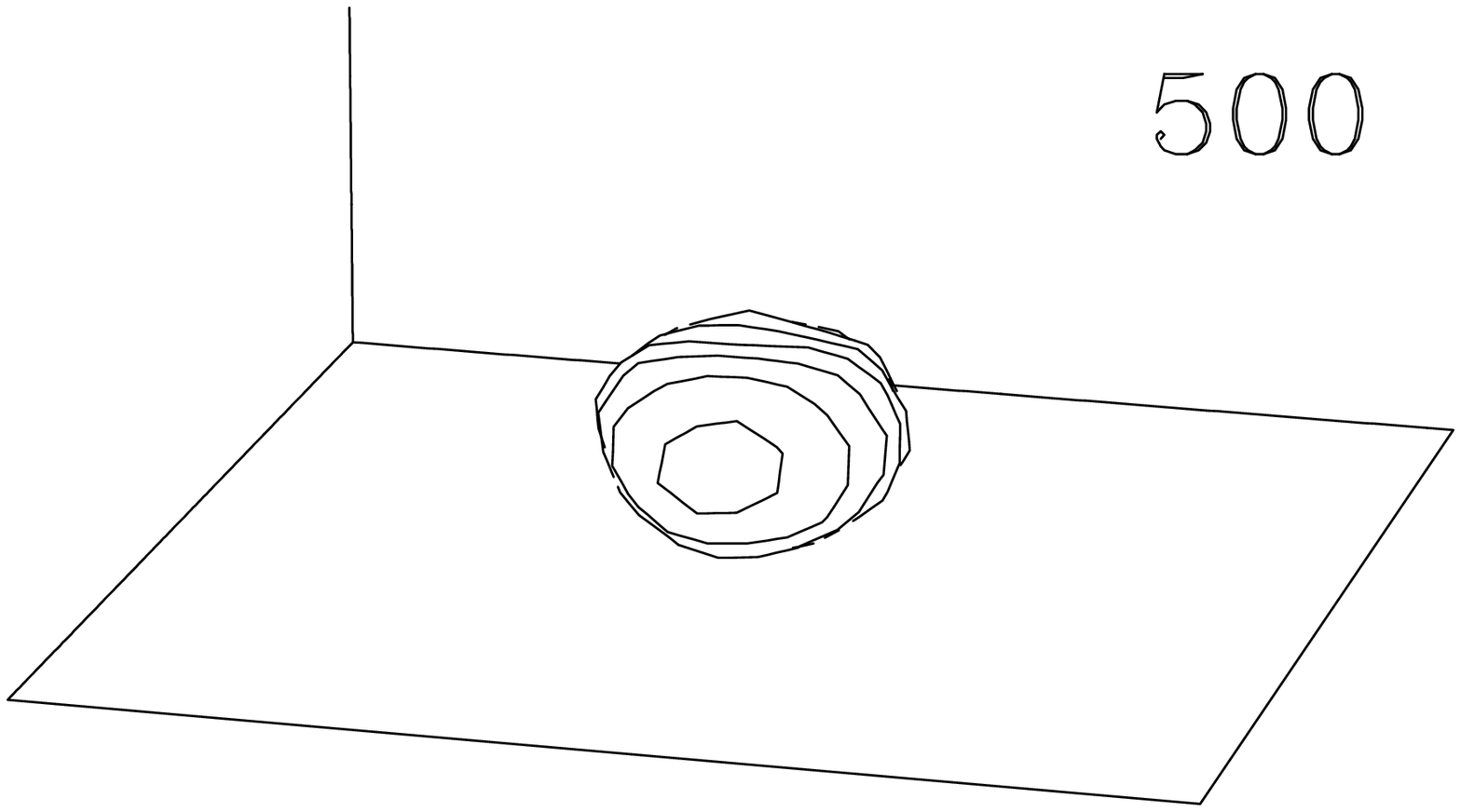}

  \vspace*{0.5in}

  \includegraphics[width=0.19\linewidth]{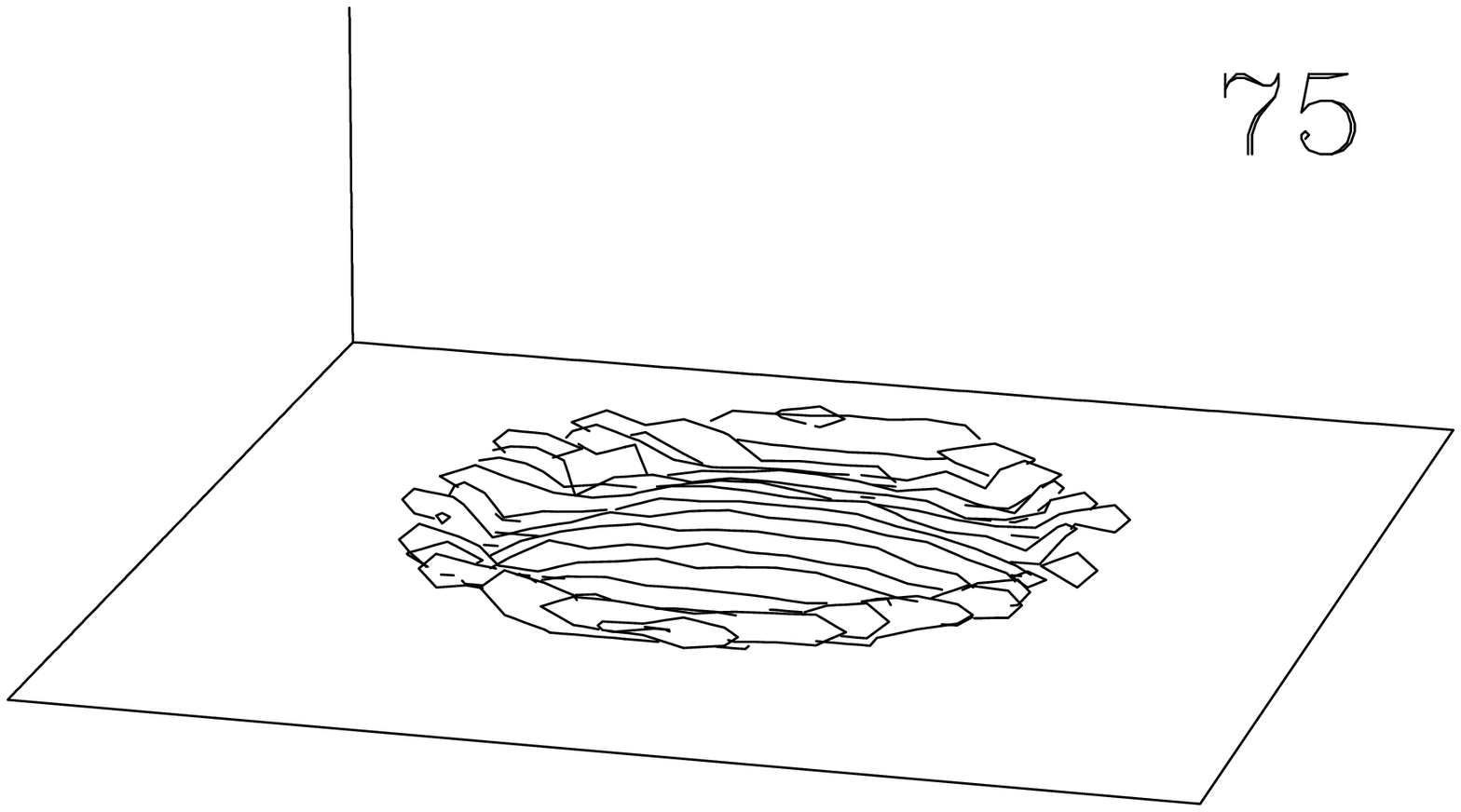}
  \includegraphics[width=0.19\linewidth]{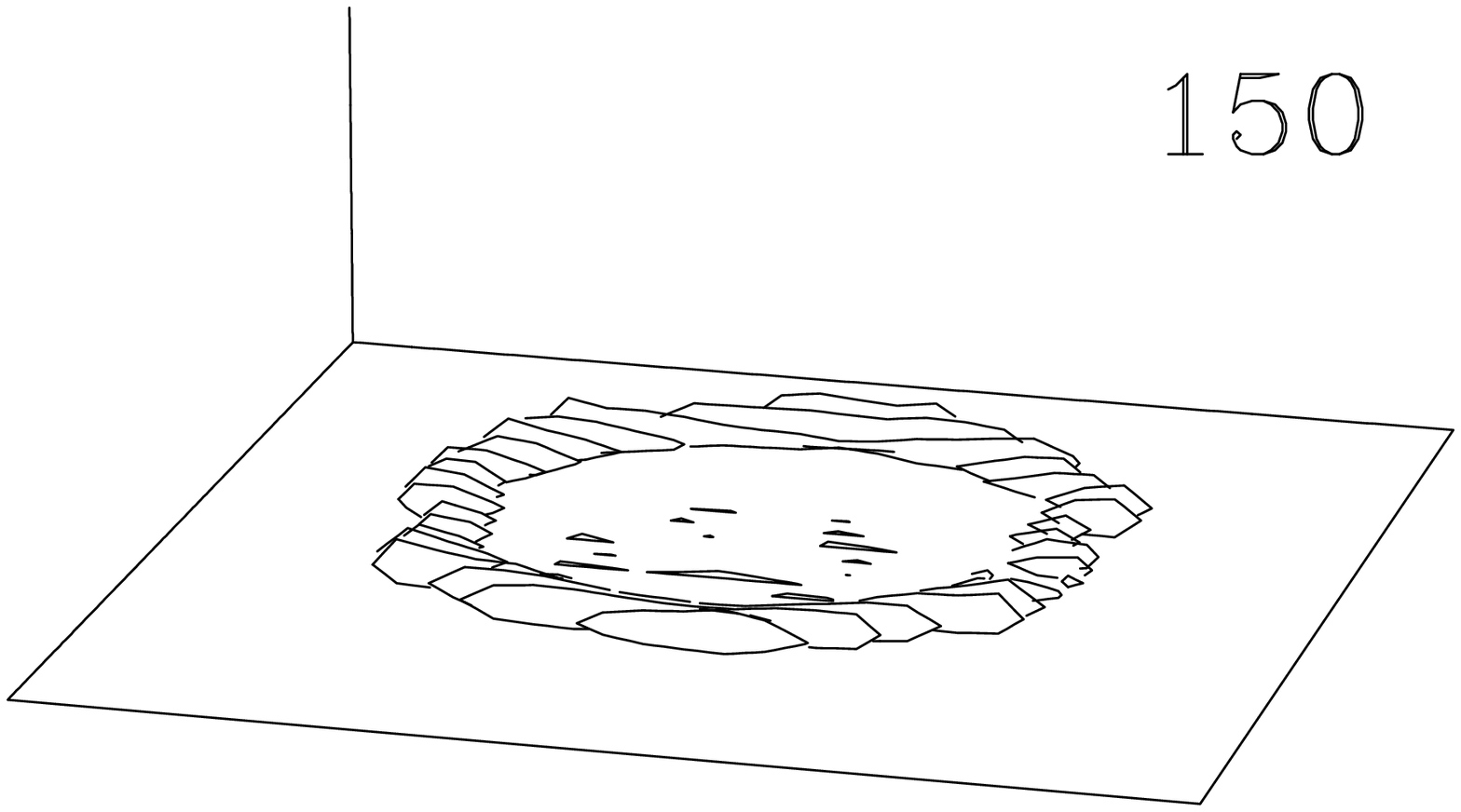}
  \includegraphics[width=0.19\linewidth]{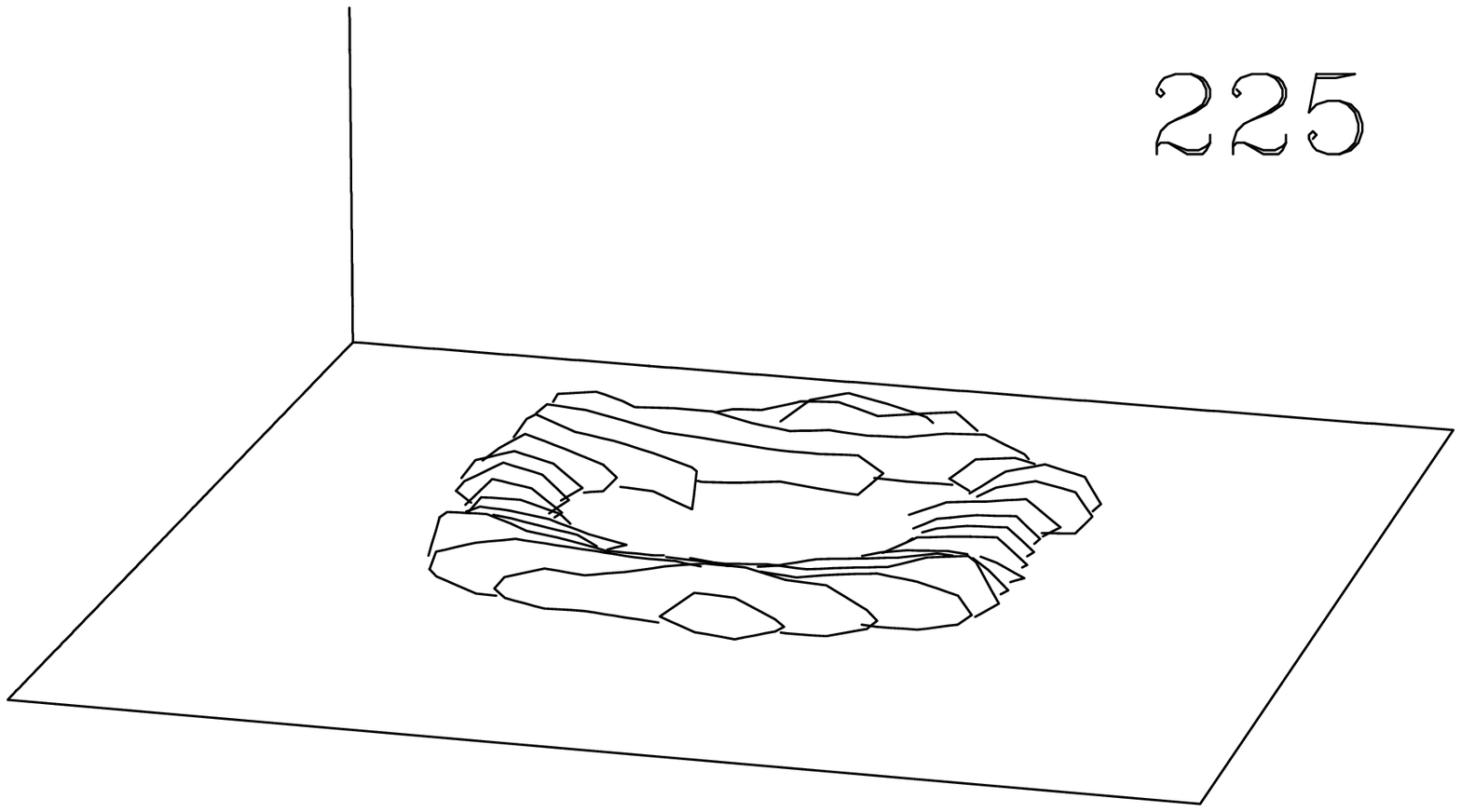}
  \includegraphics[width=0.19\linewidth]{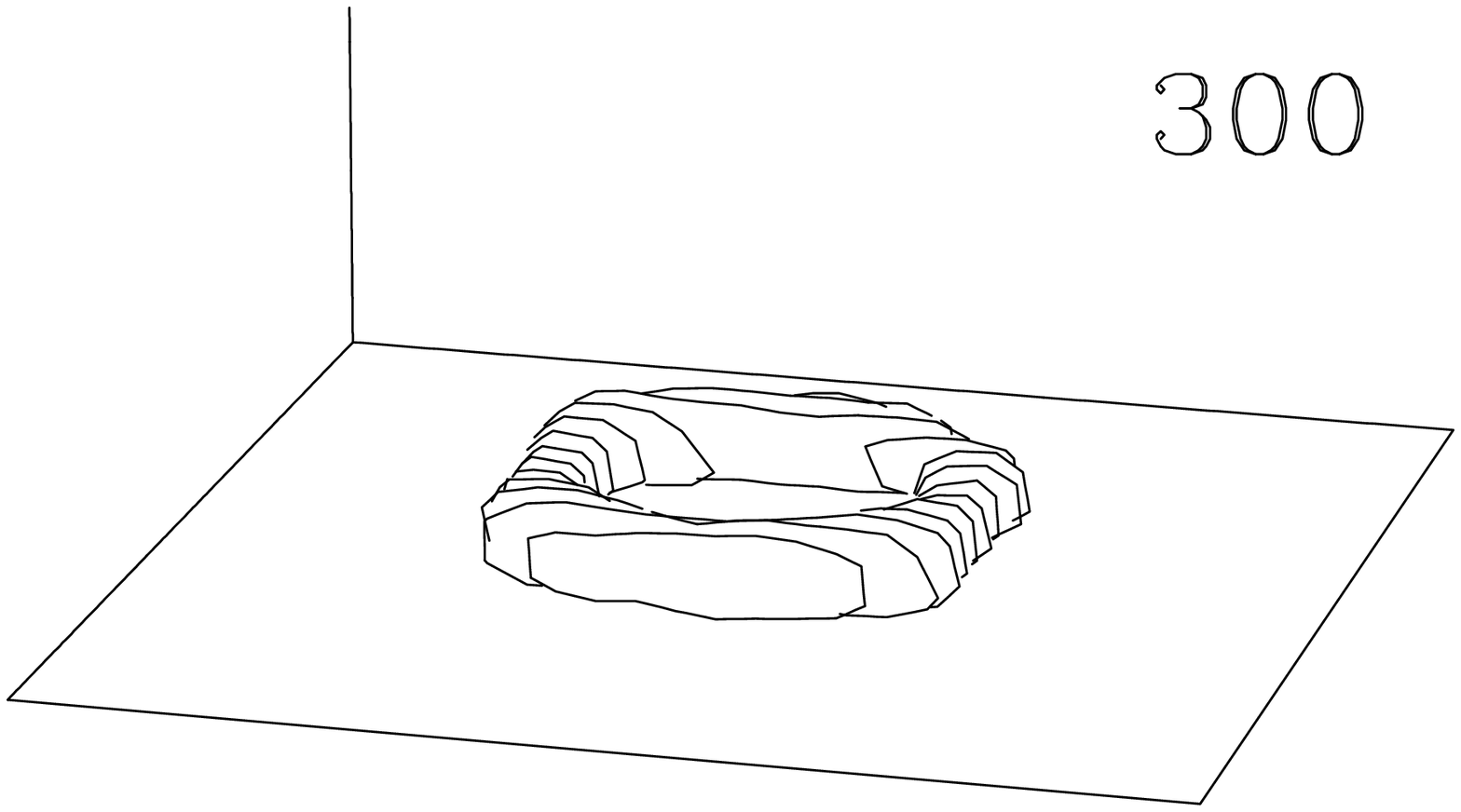}
  \includegraphics[width=0.19\linewidth]{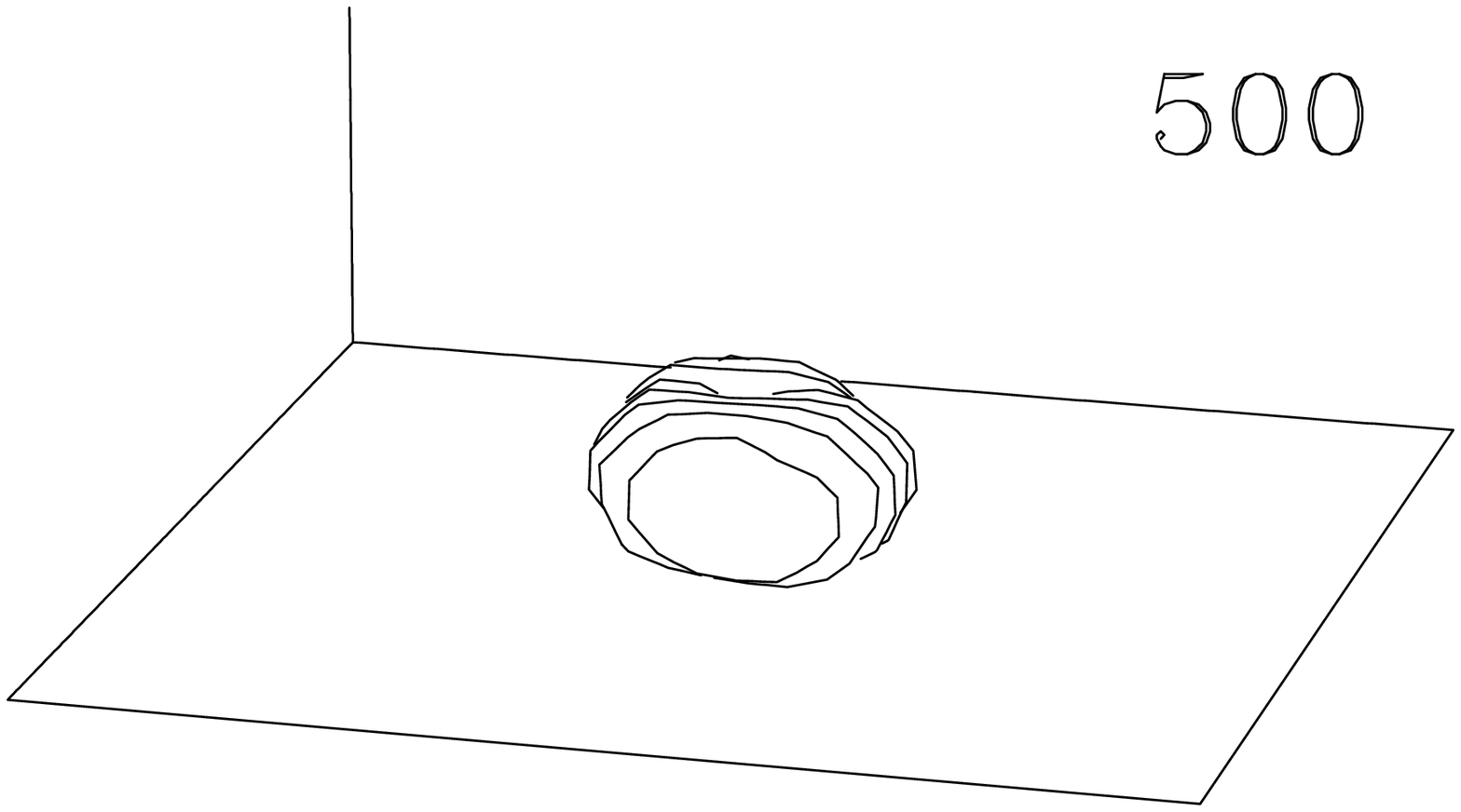}
  
  \vspace*{0.5in}

  \includegraphics[width=0.19\linewidth]{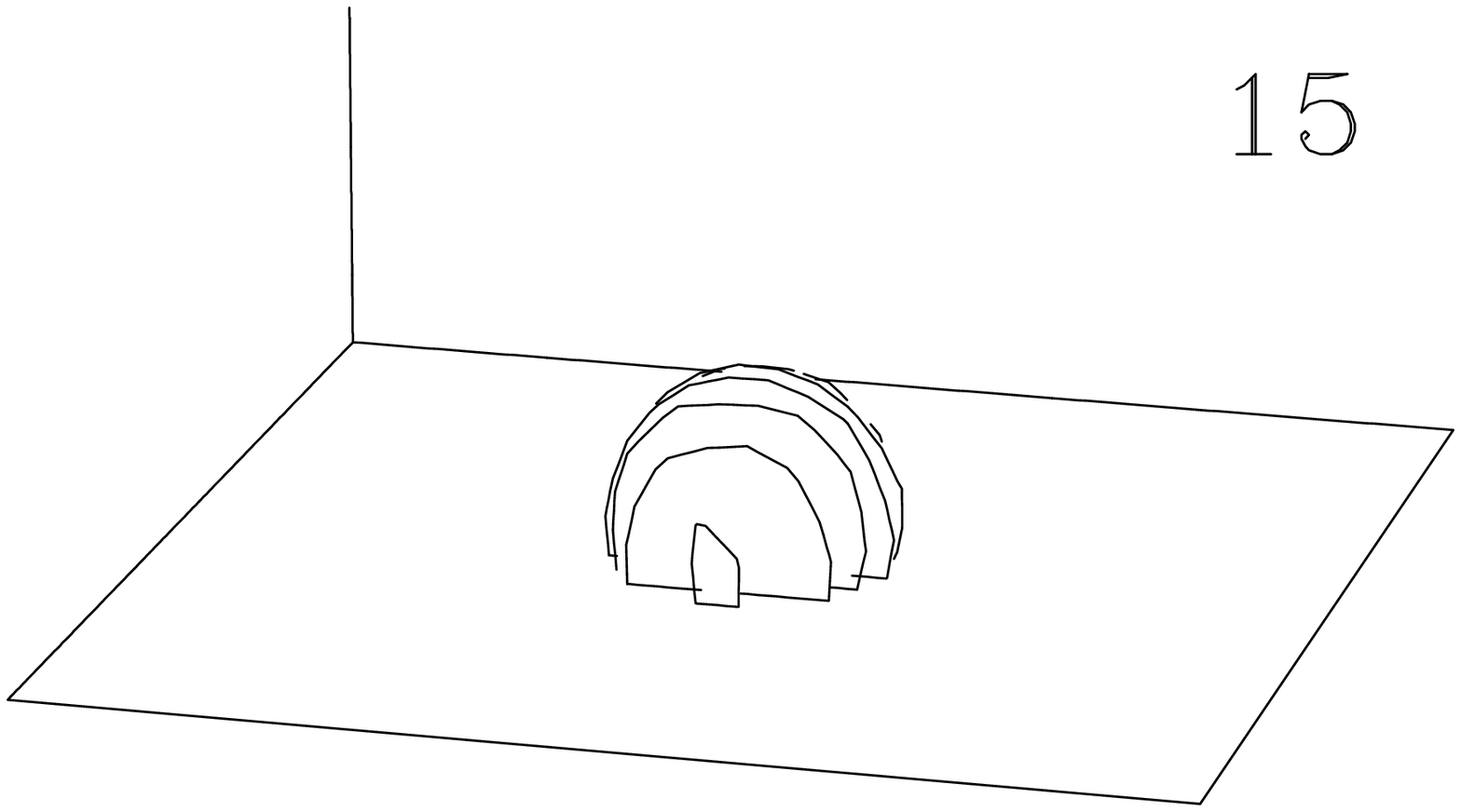}
  \includegraphics[width=0.19\linewidth]{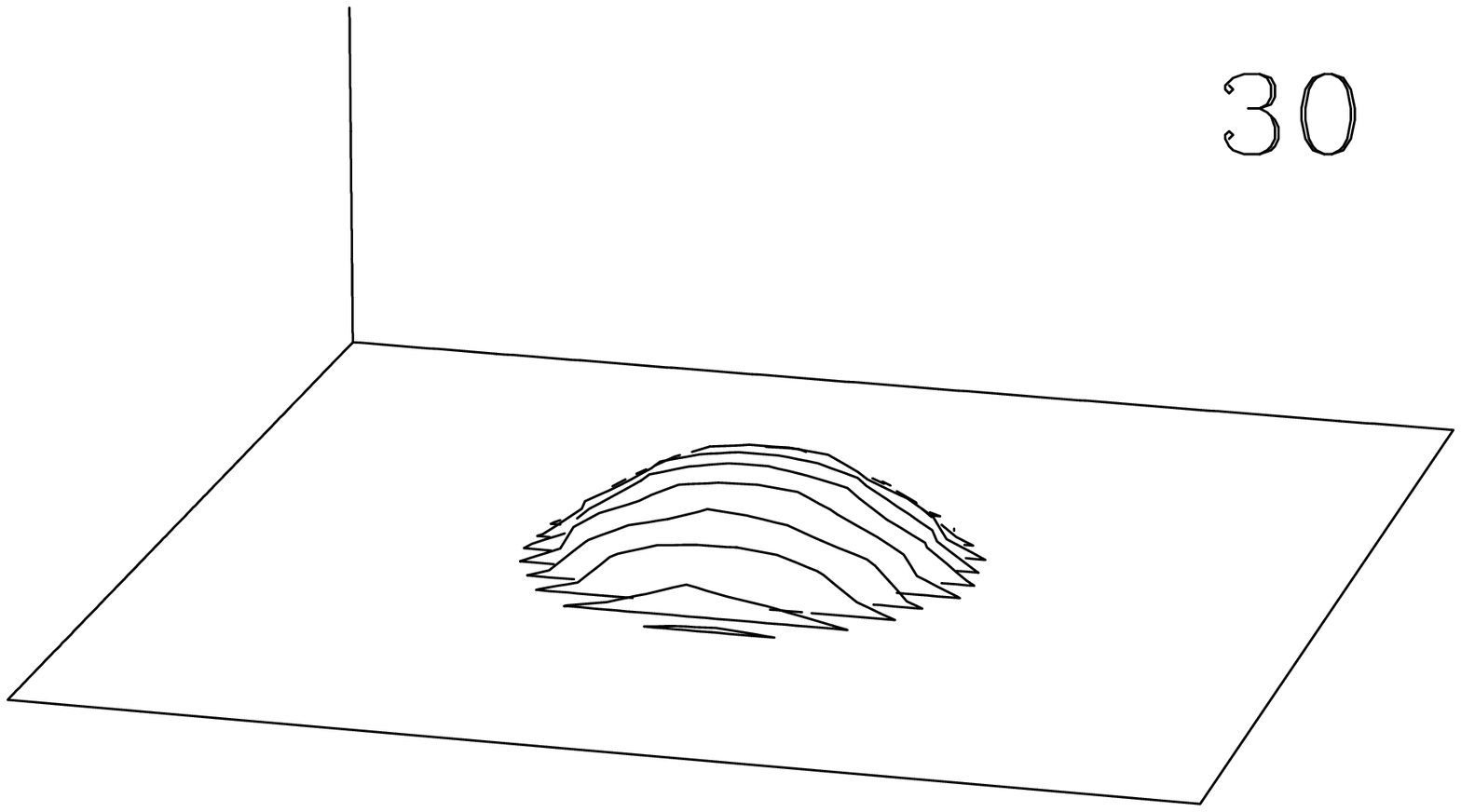}
  \includegraphics[width=0.19\linewidth]{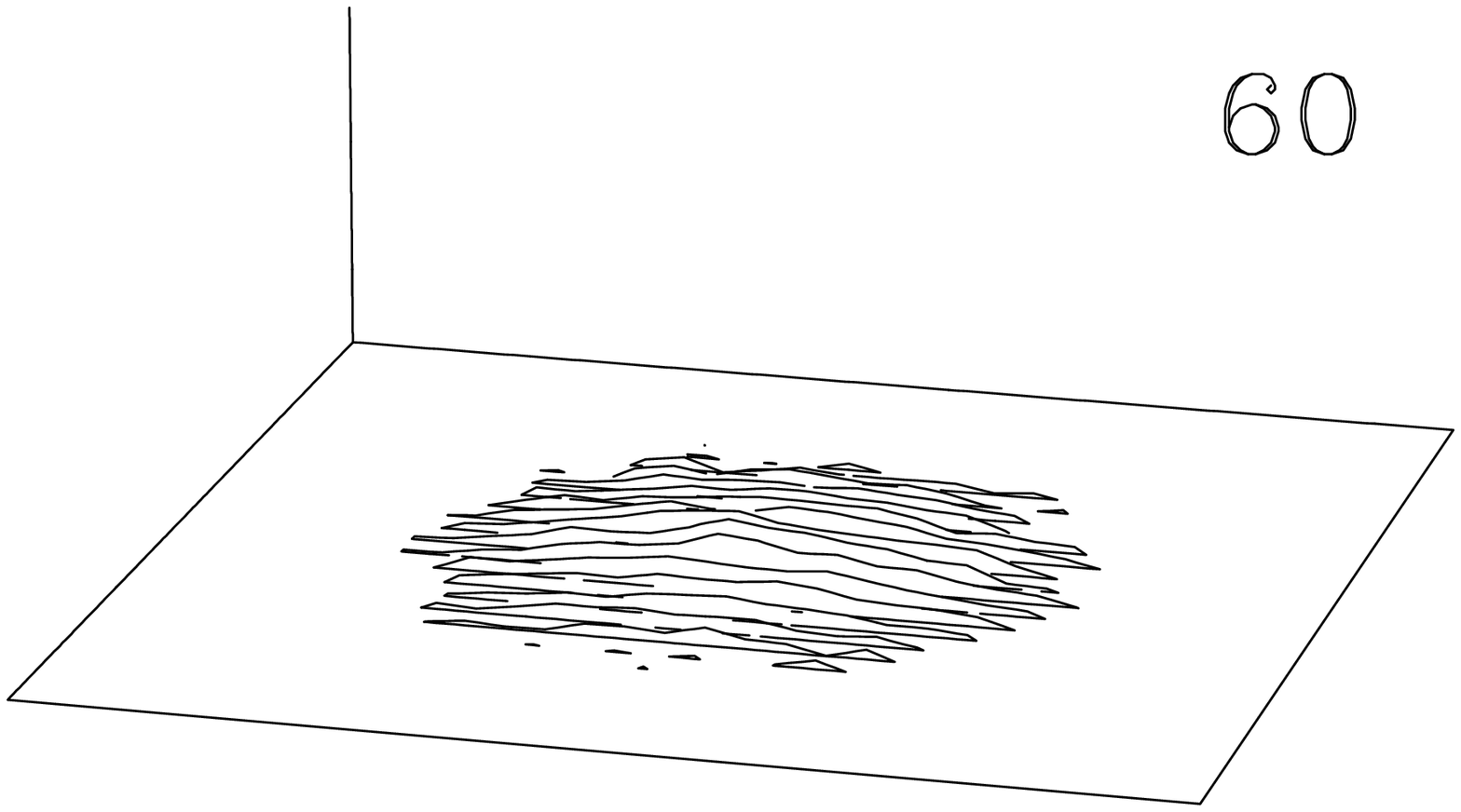}
  \includegraphics[width=0.19\linewidth]{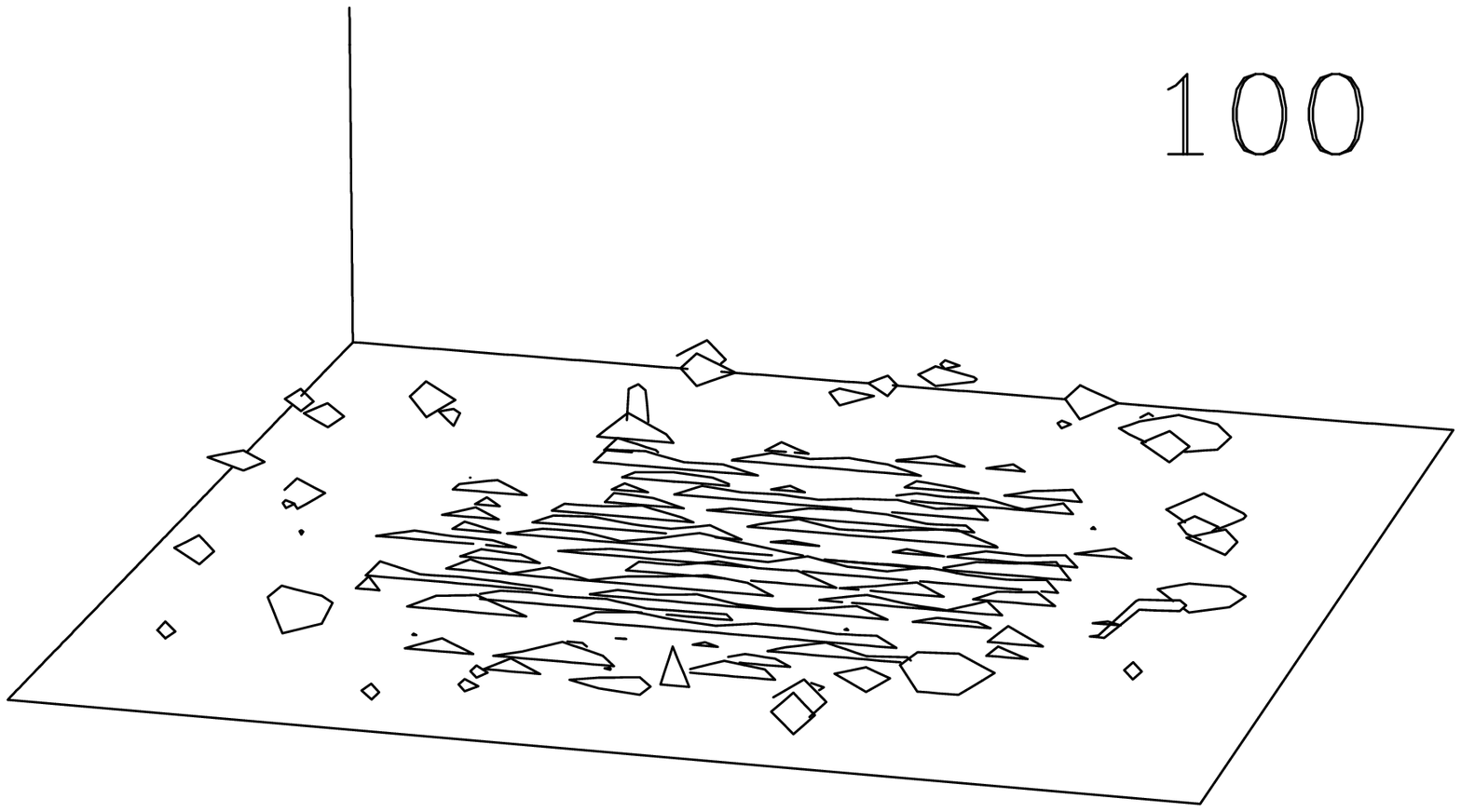}
  \includegraphics[width=0.19\linewidth]{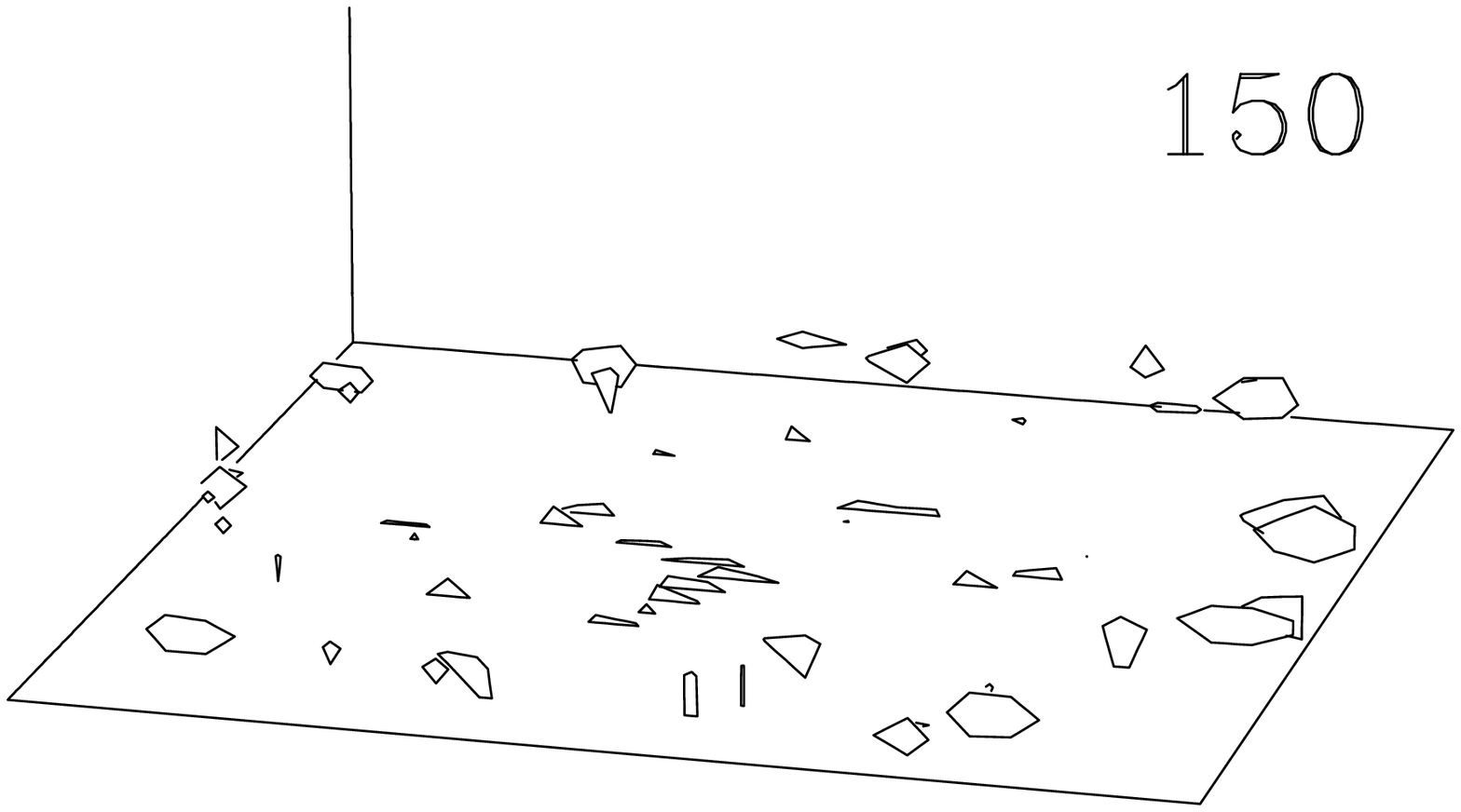}
  \caption{Snapshots of the Impact of a non-volatile drop on a non-wetting 
  solid surface, at velocities 1 (top row), 1.5 (middle row) and 
  2.0$\sigma/\tau$ (bottom row). 
  The plot depicts the mean interface of the drop at the indicated times.} 
  \label{fig1}
\end{figure}

The dimer system forms a spherical drop of radius 36$\sigma$ with an 
atomic (mass) density $\rho$=0.80$m\sigma^{-3}$, surrounded by vapor 
of atomic density 5$\times 10^{-4}m\sigma^{-3}$.  The ``edge'' of the drop is 
actually a sigmoidally-shaped transition region between liquid and vapor 
with a width of a few $\sigma$, and the ratio of vapor to liquid densities 
is about a half of that of air to water at room conditions.  The tetramer
system forms a similar drop of radius 35.8$\sigma$ and density 
$\rho$=0.86 $m\sigma^{-3}$ but in contrast to the dimers has very little vapor 
under these conditions, less than 1/10 the density 
of the dimer case.  In addition, we have in some cases also considered two 
variant systems, one 
in which a drop of tetramer is surrounded by vapor of the dimer system, as
a caricature of water drops in air, and the second a dimer system in
which the vapor molecules are removed by hand at the moment when the drop
is given its initial downward velocity.  (We do not consider a monatomic 
liquid because its vapor pressure is so high as to obscure the shape
of the interface after impact.)
The simulation box is periodic in the transverse $x$ and $z$ directions
while the bottom of the box has one layer of fcc cells of solid atoms, which
are mobile but tethered to the lattice sites by stiff springs.  The fluid 
atoms have a standard unit-strength  LJ interaction with each other, while in
most cases the interaction between fluid and solid atoms has only a 
$r^{-12}$ repulsive component, corresponding to a completely non-wetting 
system.  We occasionally refer to a completely-wetting solid where the
liquid-solid interaction has standard strength, but the effects of
wettability and patterned surfaces are explored in more detail in subsequent 
paper. The combined
liquid-vapor-solid system is equilibrated for 100$\tau$, sufficient for the
drop radii and densities to stabilize.

\begin{table}
\begin{tabular}{|l||c|c|c|} \hline
 	   			& dimer	& tetramer	\\ \hline\hline
$\:\mu$    $\;(m/\sigma\tau)$	& 2.80	& 3.42		\\ \hline
$\:\gamma$ $\;(m/\tau^2)$	& 0.51	& 0.67		\\ \hline
$\:\rho$   $\;(m\sigma^{-3})$	& 0.80	& 0.86		\\ \hline
$\:R$	   $\;(\sigma)$		& 36.0  & 35.8		\\ \hline
\end{tabular}
\label{table1}
\caption{Physical properties of the drops.}

\vspace*{0.25in}

\begin{tabular}{|c||c|c||c|c|} \hline
$u_0$	& \multicolumn{2}{c|}{dimer} & \multicolumn{2}{c|}{tetramer} \\
\hline\hline
	& $Re$	& $W\!e$& $Re$  & $W\!e$\\ \hline
1.0	& 10.3	& 31.	& 9.0	& 46.0  \\ \hline
1.5	& 15.5	& 127.	& 13.5	& 104.  \\ \hline
2.0	& 41.2	& 226.	& 18.0	& 184.  \\ \hline	
3.0	& 31.0	& 552.	& 27.0	& 208.  \\ \hline	
\end{tabular}
\label{table2}
\caption{Simulation parameters}
\end{table}

After equilibration the velocity of each atom in the drop is shifted by a
negative constant in the vertical direction so as to translate the drop 
downward normally towards the solid surface.  
In the volatile case this additional velocity is smoothly turned off outside 
the interfacial region and the external vapor velocities are not shifted 
at all.  In the remainder of the simulation the solid surface is held at
constant temperature, but the temperature of the drop and vapor are allowed to
vary.  In Table I we give the viscosity $\mu$ and surface tension $\gamma$ 
of the dimer and tetramer liquids used here, at the initial temperature of 
the simulations, along with the drop densities and radii.  The viscosities
and surface tensions 
were determined by standard methods \cite{arfm}, using independent simulations 
of Couette flow, and analysis of a periodic slab of liquid in contact with 
vapor, respectively.  In Table II we list the initial impact velocities $u_0$ 
and the corresponding values of the 
Reynolds and Weber numbers, $Re=\rho u_0 R/\mu$ and $W\!e=\rho u_0^2
R/\gamma$.  We have also simulated both lower (0.3) and higher (10.0)
initial velocity impacts, but these do not exhibit any qualitatively
different behavior.  Given the fixed properties of the
two liquids, the only remaining variable parameter is the initial velocity
and $W\!e/Re^2$ is a material constant.  

Although the values of $Re$ and
$W\!e$ fall in a range suitable for comparison to laboratory experiments, one
distinctly different property is the Mach number, $Ma=u_0/u_s$ where $u_s$
is the sound speed.  In the Appendix we present measurements of the sound
speed in the two liquids. which gives an estimate $u_s \sim 5 \sigma/\tau$.
As a result the Mach number in these simulations is not negligible, ranging 
from 0.2 -- 0.5.  The underlying reason for
high $Ma$ in comparison to typical laboratory systems is that 
it is necessary to use relatively large velocities (typically 0.01 - 1.0
$\sigma/\tau$) in molecular simulations to see any signal over the thermal 
fluctuation noise, whereas the thermal velocity is O(1) in these units.
However, we find that significant density variations appear only at the
highest velocities studied here. 

\begin{figure}[t]
  \includegraphics[width=0.19\linewidth]{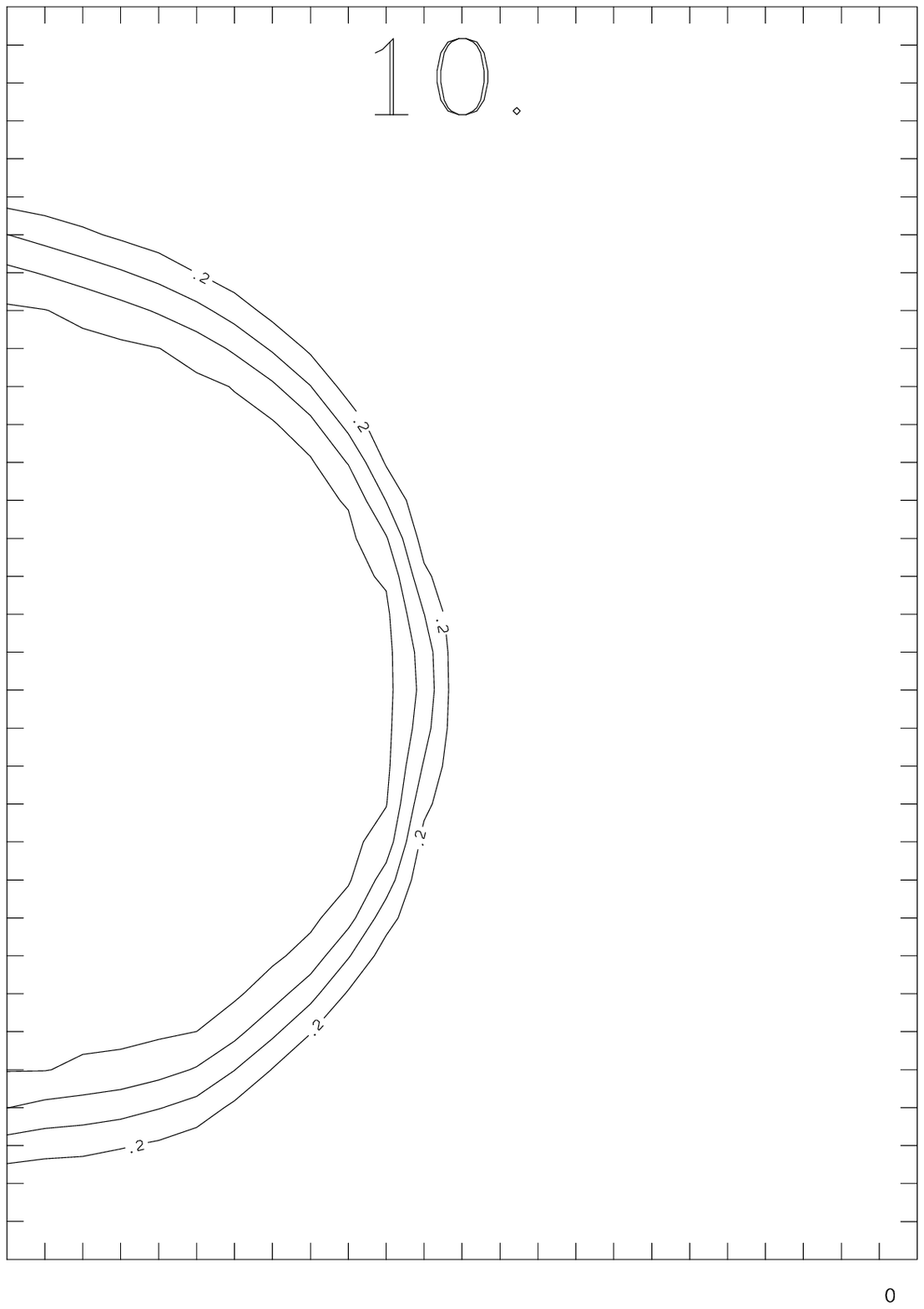}
  \includegraphics[width=0.19\linewidth]{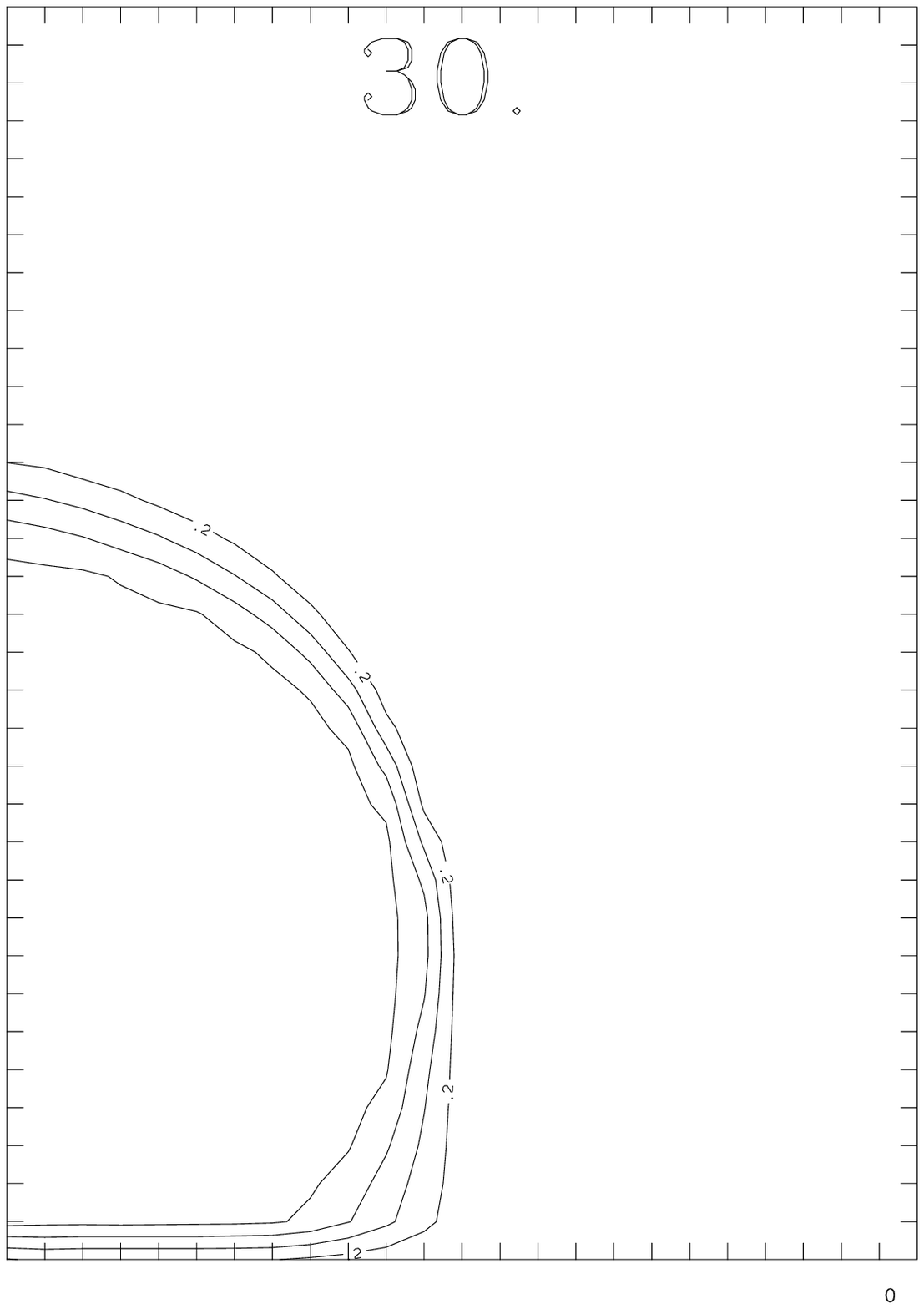}
  \includegraphics[width=0.19\linewidth]{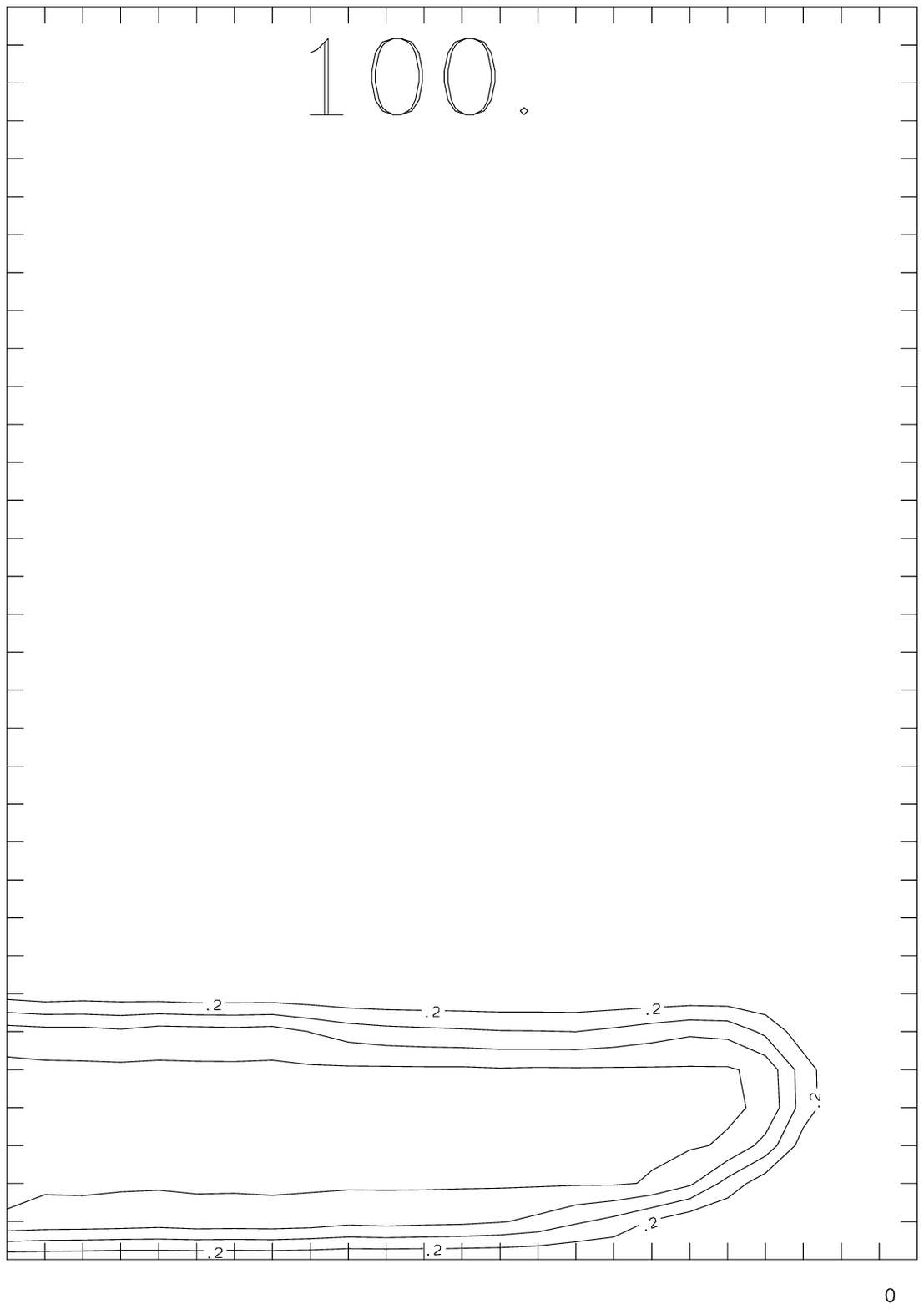}
  \includegraphics[width=0.19\linewidth]{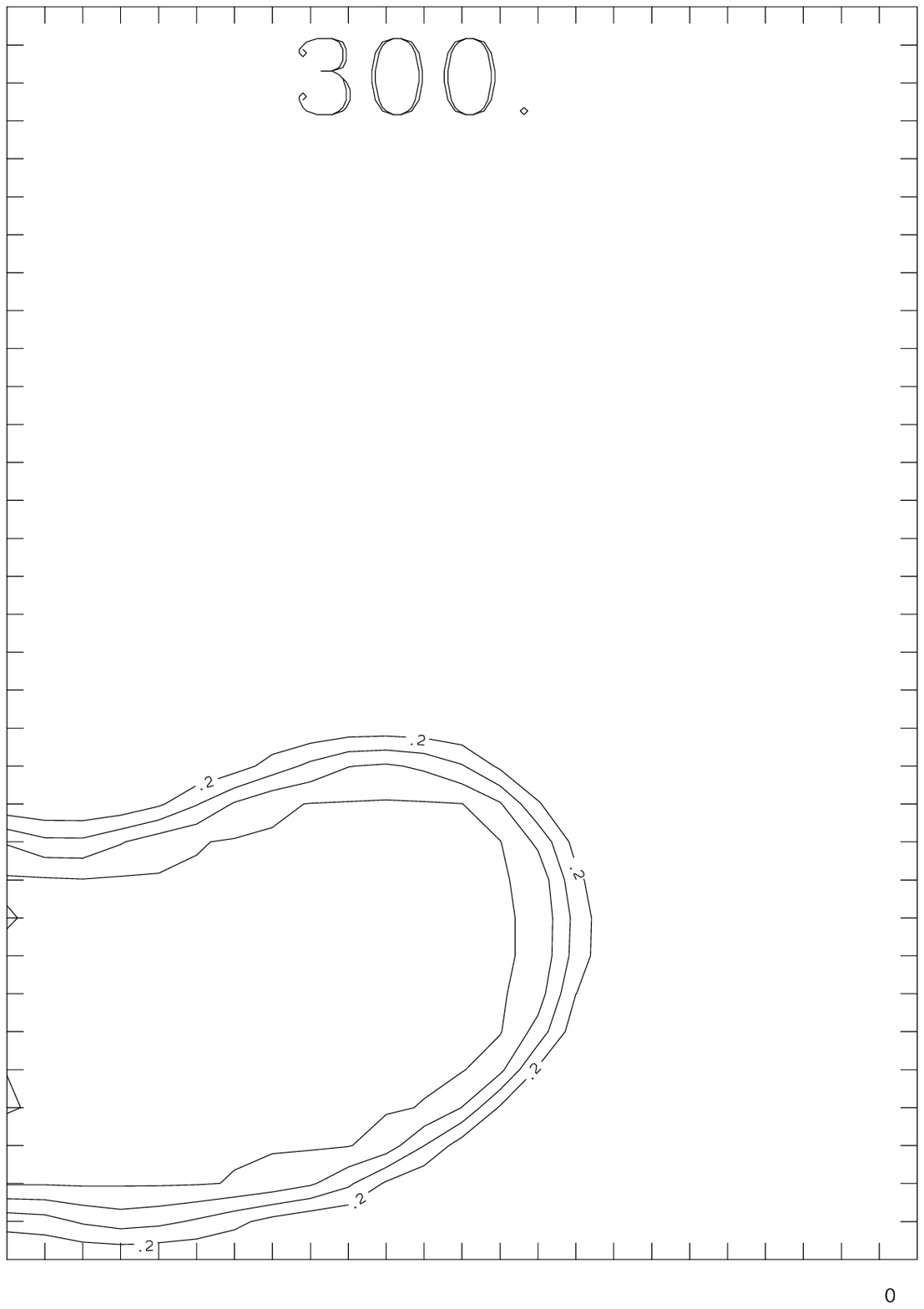}
  \includegraphics[width=0.19\linewidth]{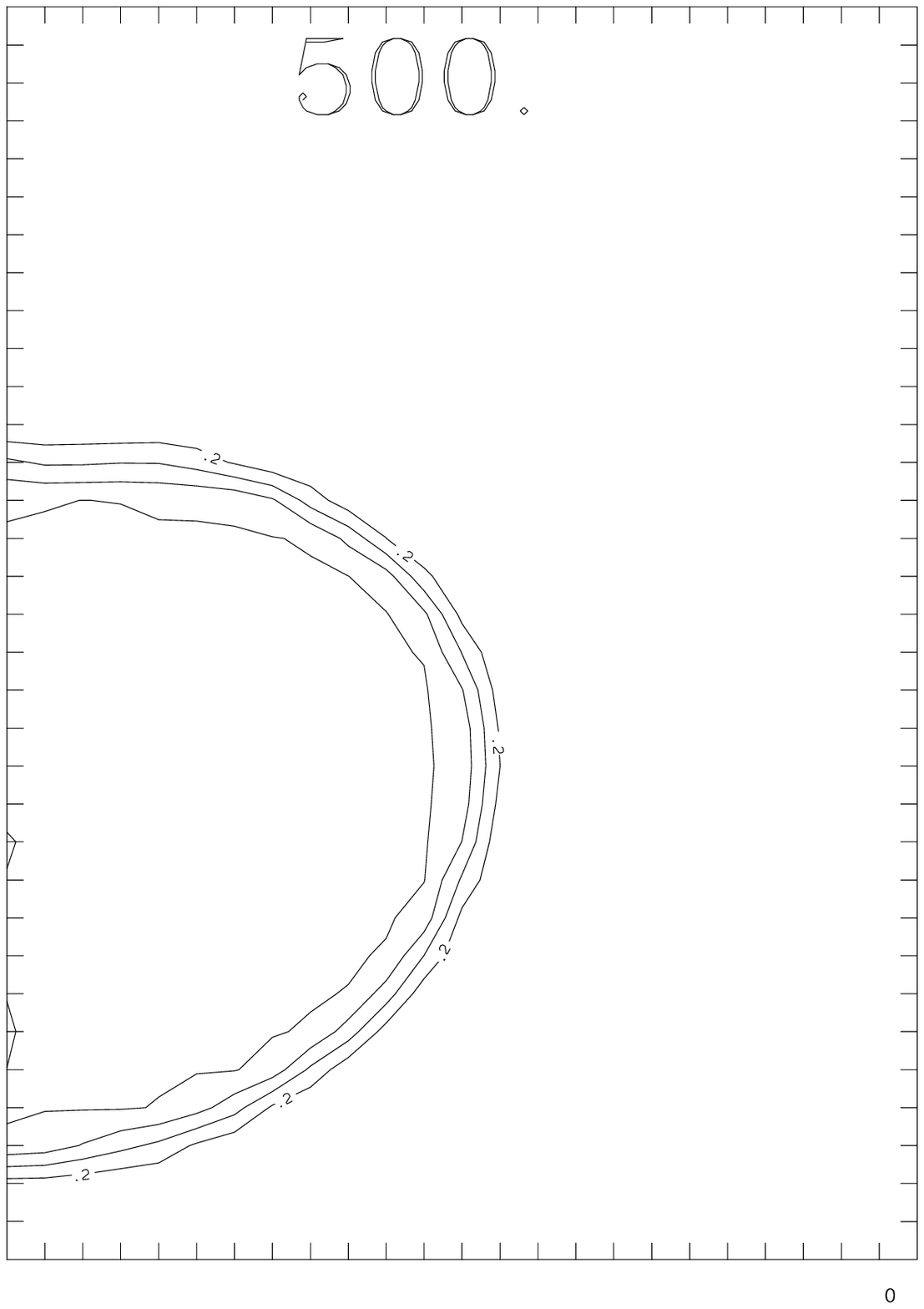}

  \vspace*{0.5in}

  \includegraphics[width=0.24\linewidth]{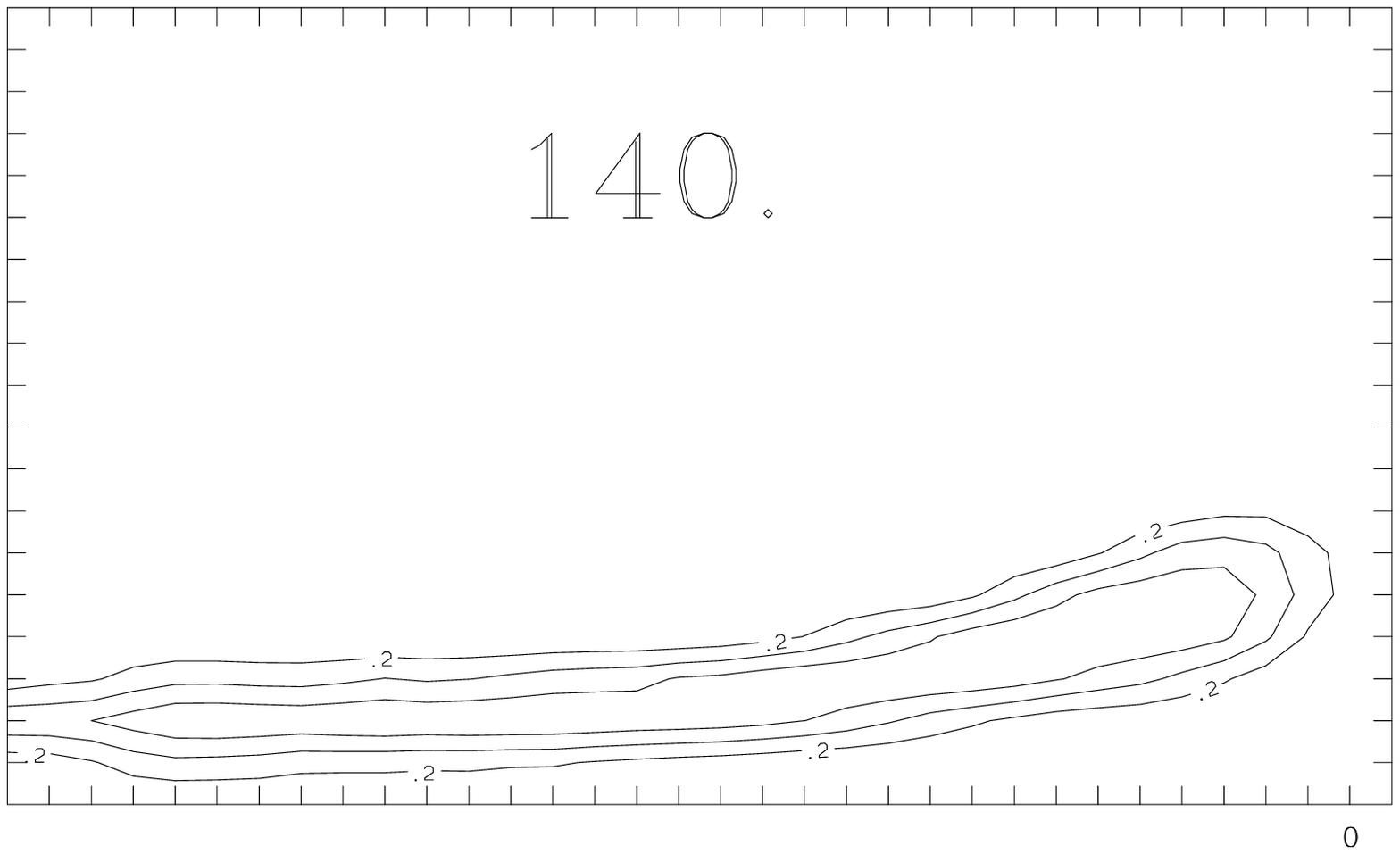}
  \includegraphics[width=0.24\linewidth]{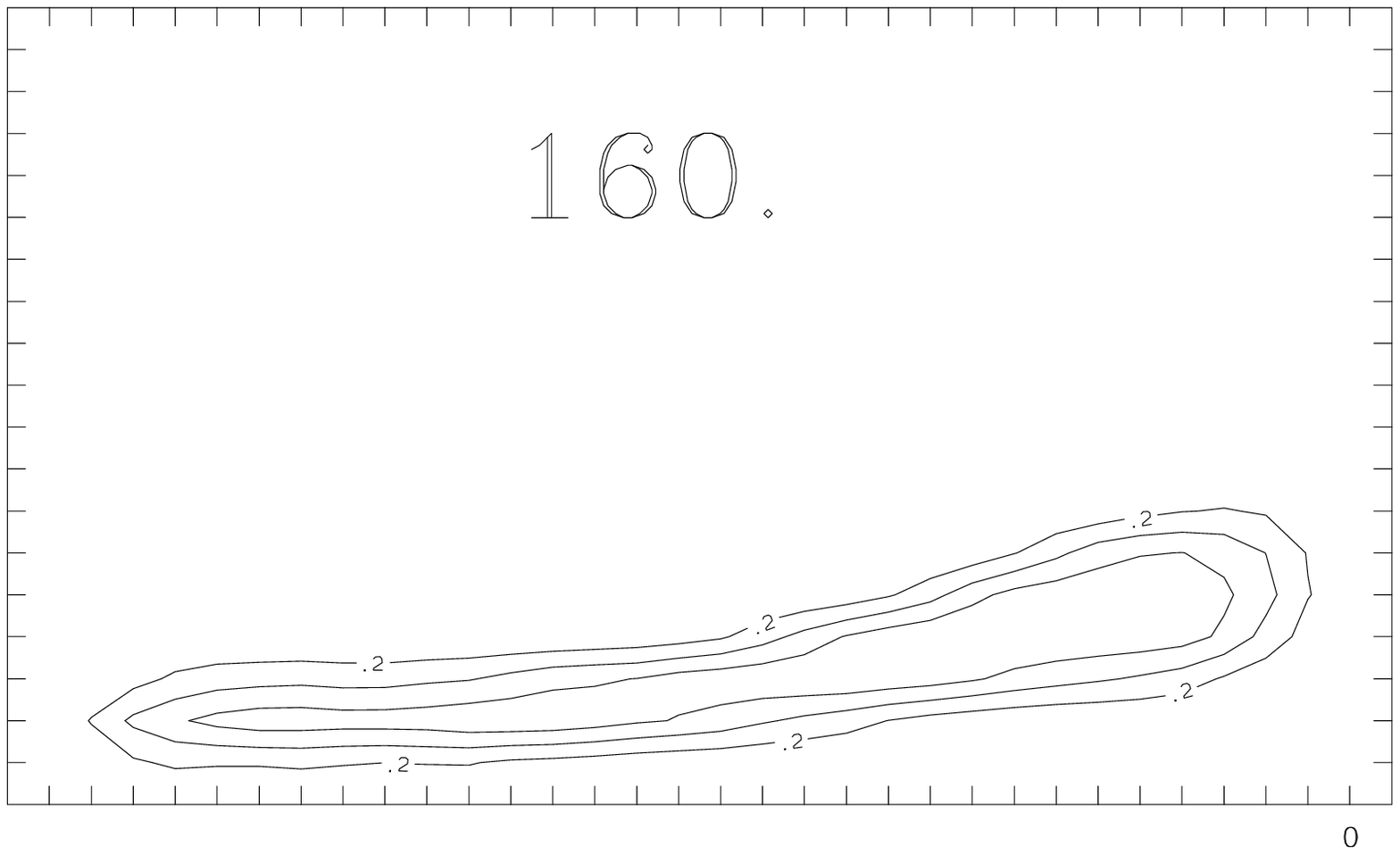}
  \includegraphics[width=0.24\linewidth]{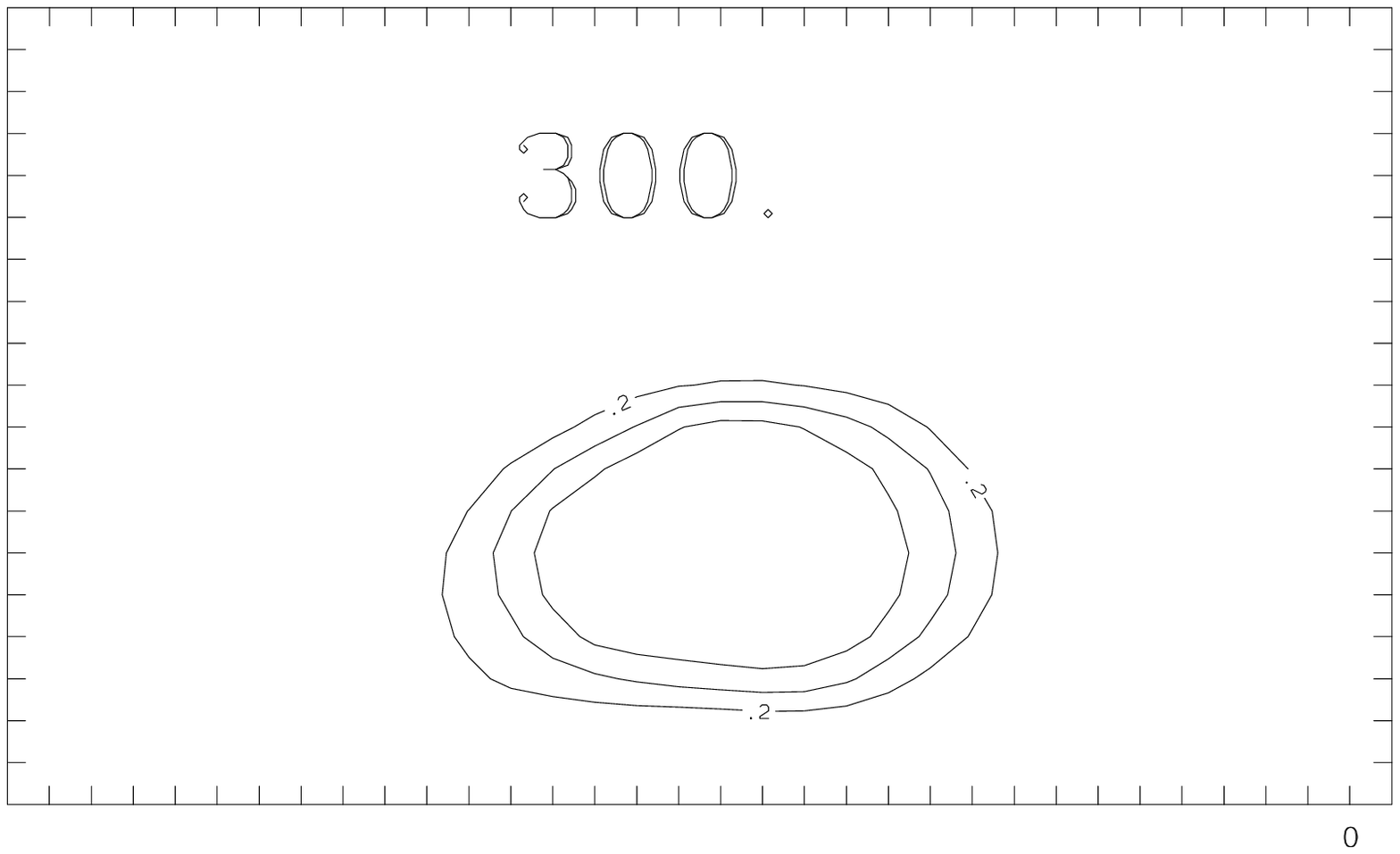}
  \includegraphics[width=0.24\linewidth]{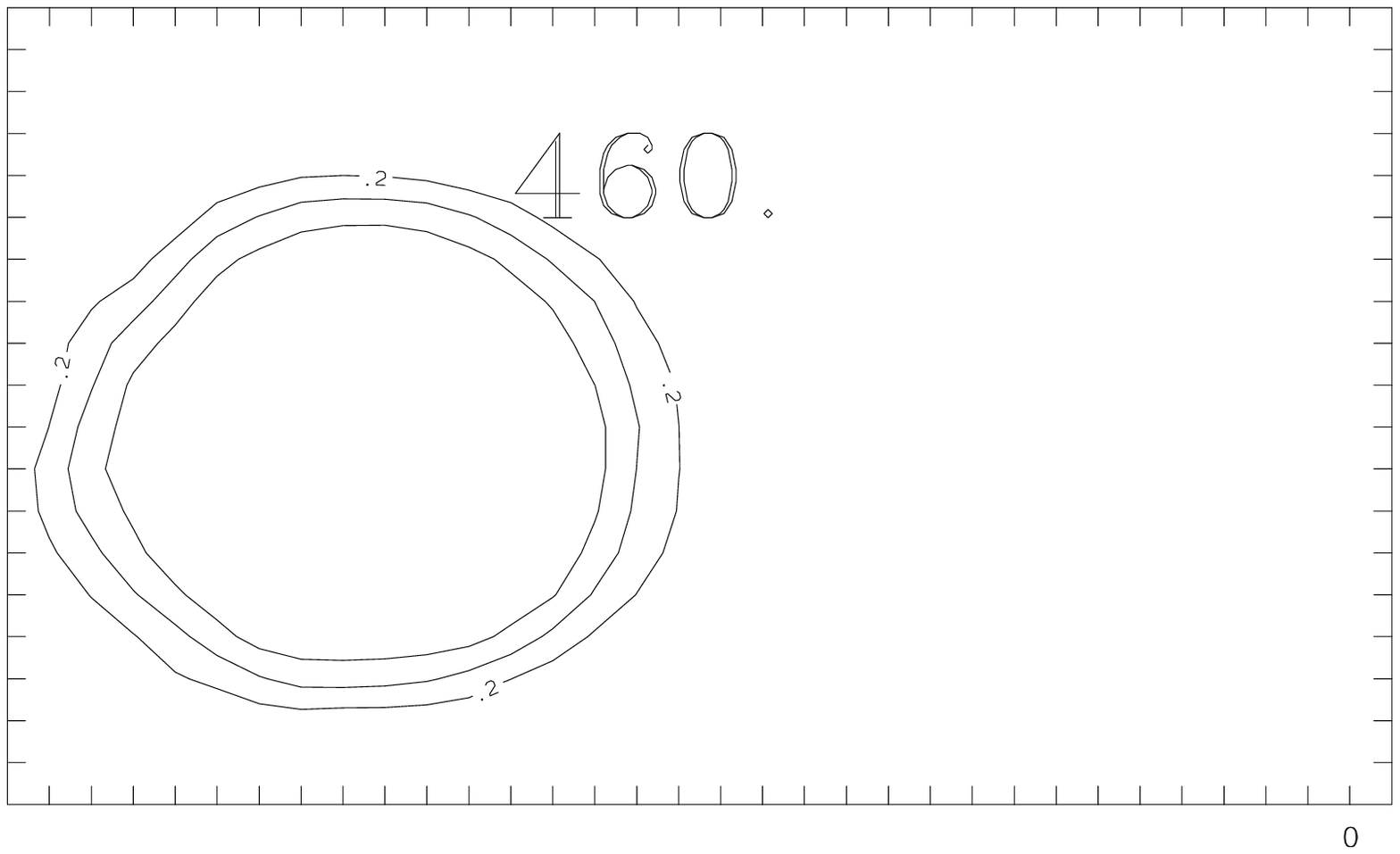}

  \vspace*{0.5in}

  \includegraphics[width=0.32\linewidth]{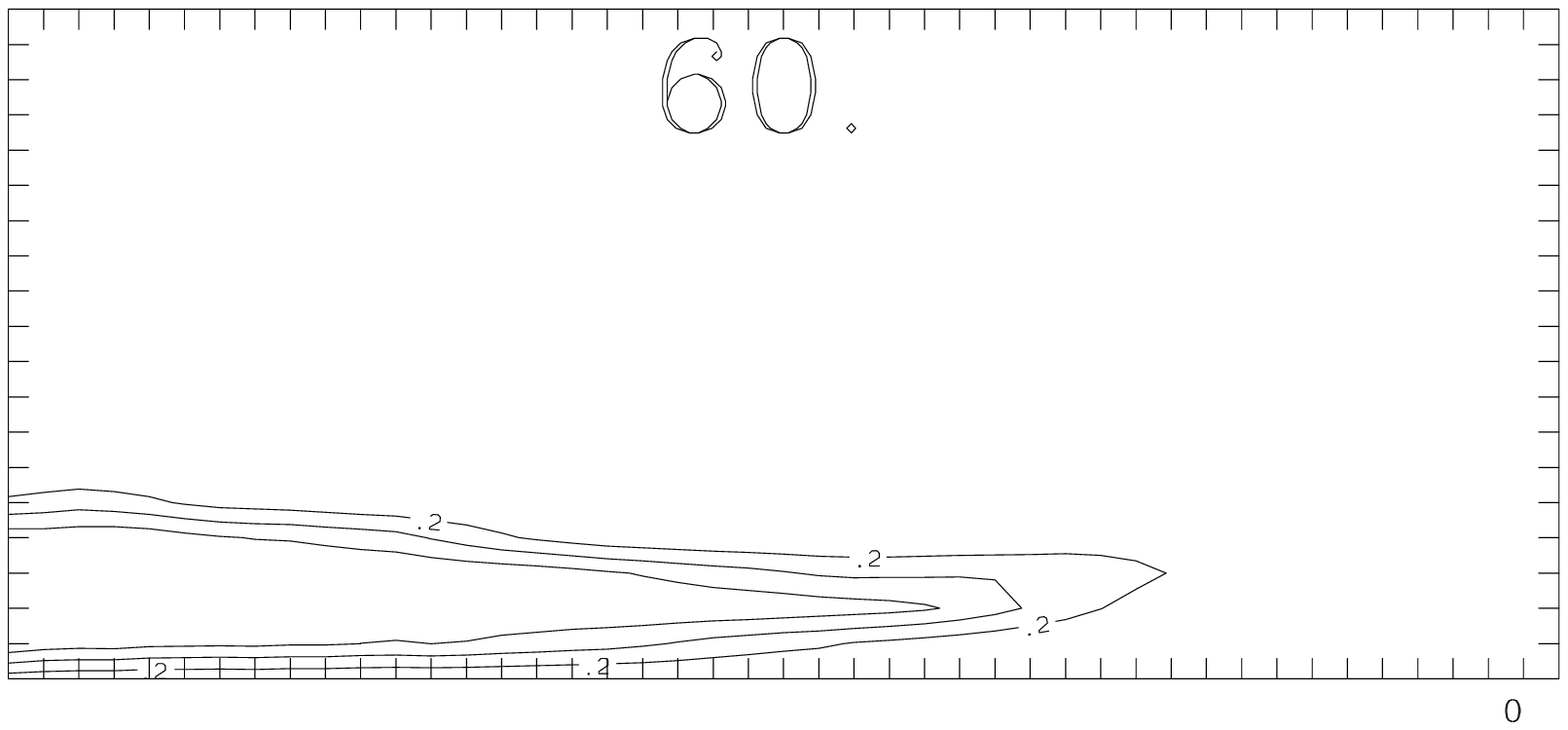}
  \includegraphics[width=0.32\linewidth]{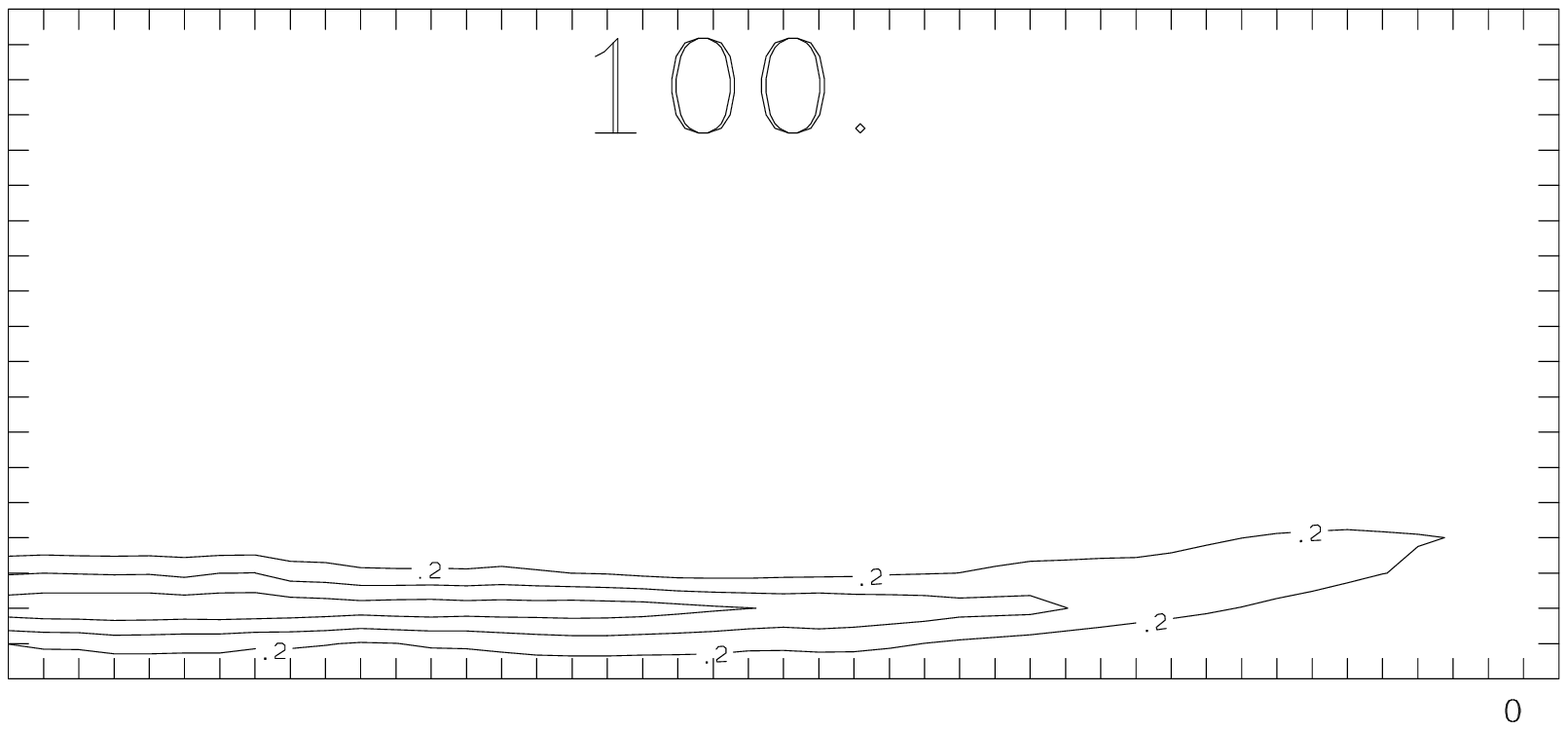}
  \includegraphics[width=0.32\linewidth]{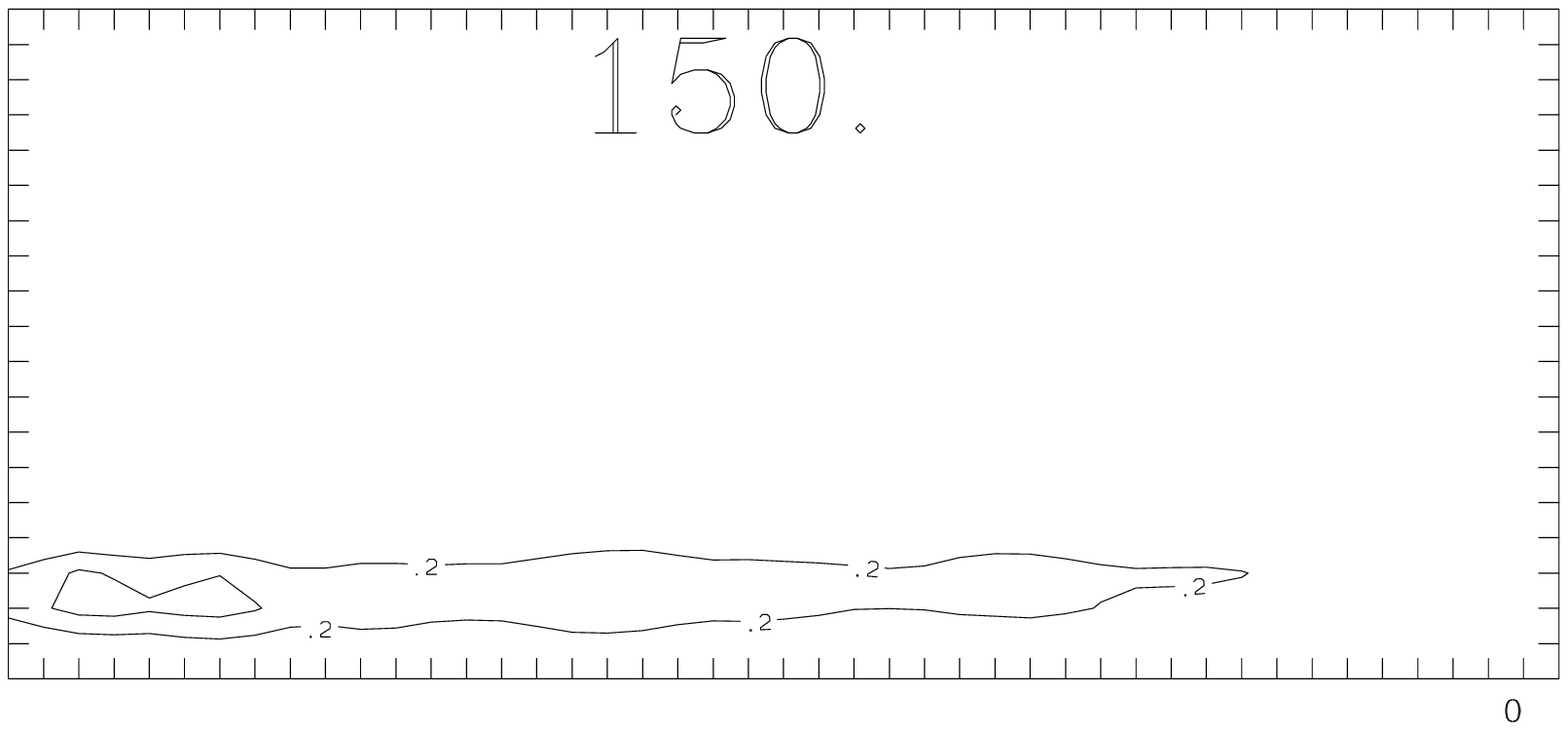}
  \caption{Density contours for impacting drops on non-wetting surfaces, 
  averaged over the 10$\tau$ 
  interval preceding the indicated time, in cylindrical coordinates
  relative to the impact axis. The initial density at the center of the drop is
  0.86 $m\sigma^{-3}$, the outer contour in each frame represents density 0.2, 
  and the inner contours are spaced by 0.2. Top to bottom:
  impact velocities 1.0, 1.5 and 2.0 $\sigma/\tau$.}
  \label{fig2}
\end{figure}

\section{results}
\label{sec:results}
\subsection{Shape and density}

We first consider the non-volatile tetramer liquid impacting a non-wetting
surface at different speeds.  
In Fig.~\ref{fig1} we show the evolution of the drop shape in terms of
the ``mean liquid-vapor interface,'' the surface on which the fluid density 
is half of that in the original drop interior.  In all cases, the drop is 
initially spherical and after contact continues to fall vertically and
initially distorts into a hemisphere.
At impact velocity 1.0 $\sigma/\tau$ the drop bounces, first spreading
radially into a flat-topped lamella or pancake with maximum extension at about
100$\tau$, and then curling up at the edges and beginning to withdraw while
remaining in contact with the surface. Subsequently the drop lifts off the
surface and continues to contract into a sphere, although the completion of
the latter
process requires several hundred $\tau$ beyond the last frame shown.  
The shape evolution of the drops is clearest in terms of
density contours:  using the centerline of the falling drop as the axis of a 
fixed cylindrical coordinate system, we average over azimuthal angle and 
measure the density of atoms as a function of height and radius, averaged over 
10$\tau$ intervals. A selection of the resulting density contours 
corresponding to the surface plots at the same three impact velocities 
are given in Fig.~\ref{fig2}. In the low-velocity bounce, the contact angle 
first decreases from 180$^\circ$ at contact to to 90$^\circ$ at 30$\tau$
as seen previously, to roughly 45$^\circ$ at time 50$\tau$ (not shown). 
Afterwards the rim of the lamella rises and the angle increases as
the drop contracts.  The curvature at the top of the drop is initially
positive and decreases until the top is roughly flat at maximum
lateral extension, then goes negative as the drop's edges  contract and
rise, and eventually returns to positive as the drop rises and slowly
returns to a sphere.  

\begin{figure}[t]
  \begin{center}
  \includegraphics[width=0.3\linewidth]{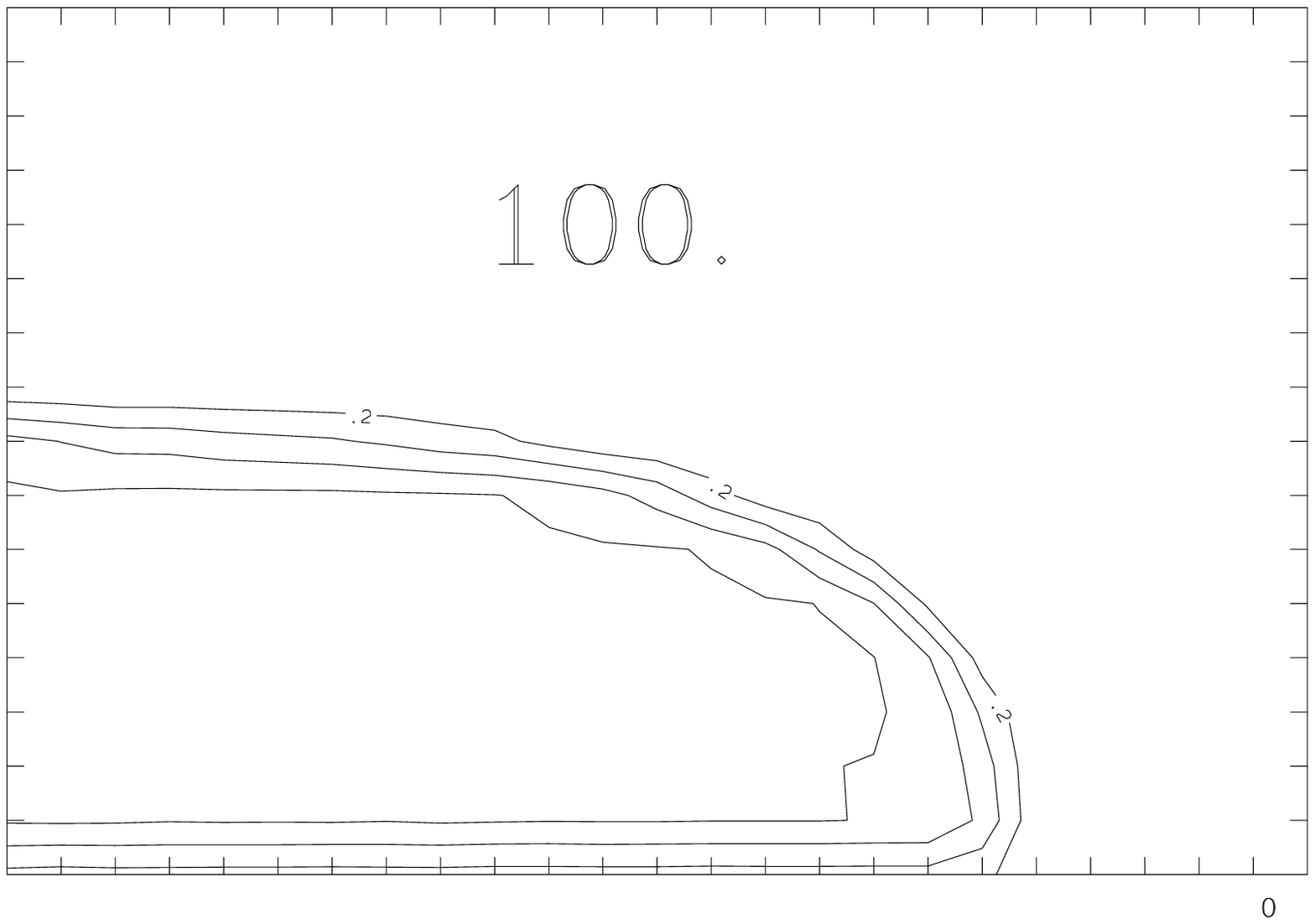}\hspace{0.1in}
  \includegraphics[width=0.3\linewidth]{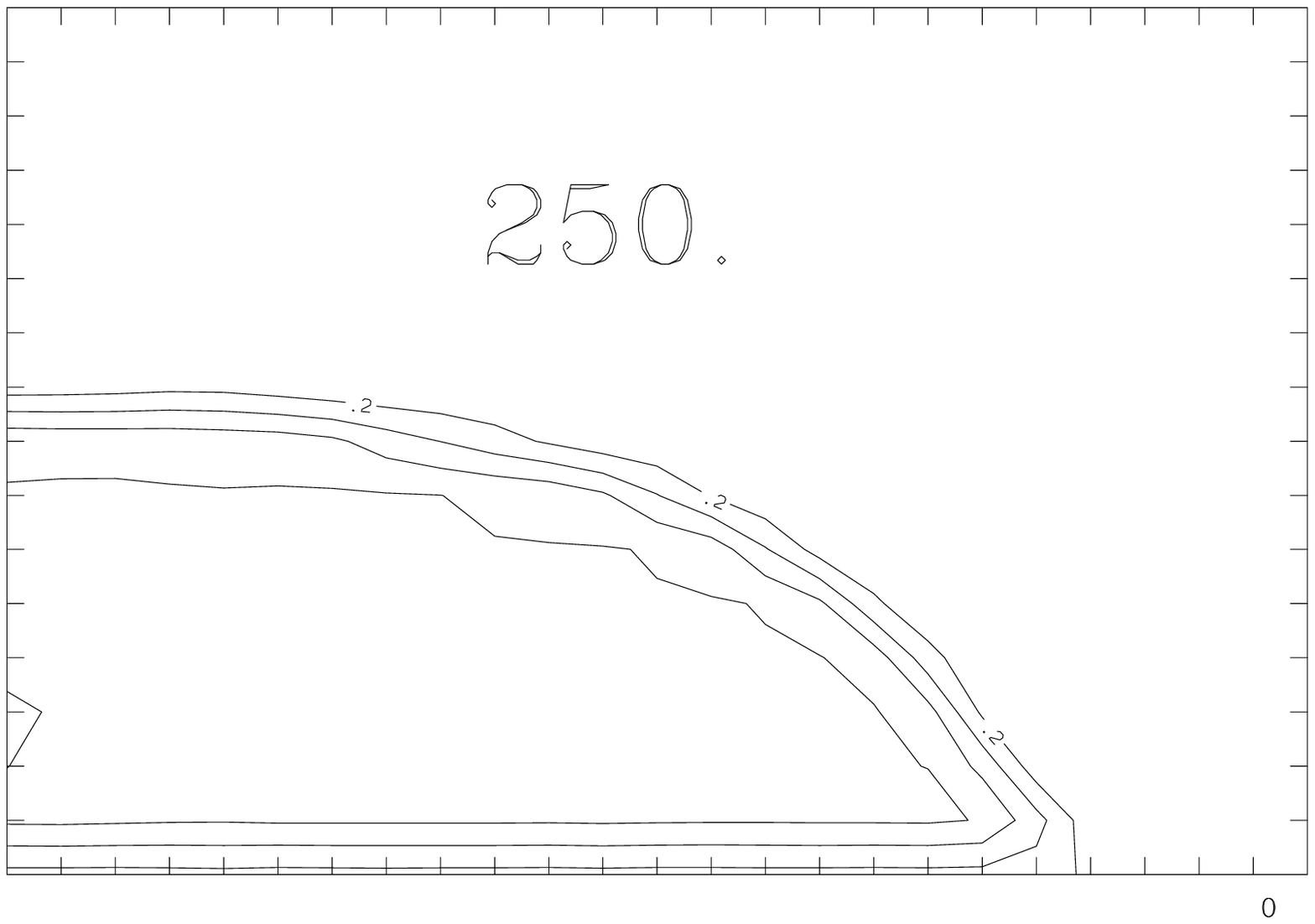}\hspace{0.1in}
  \includegraphics[width=0.3\linewidth]{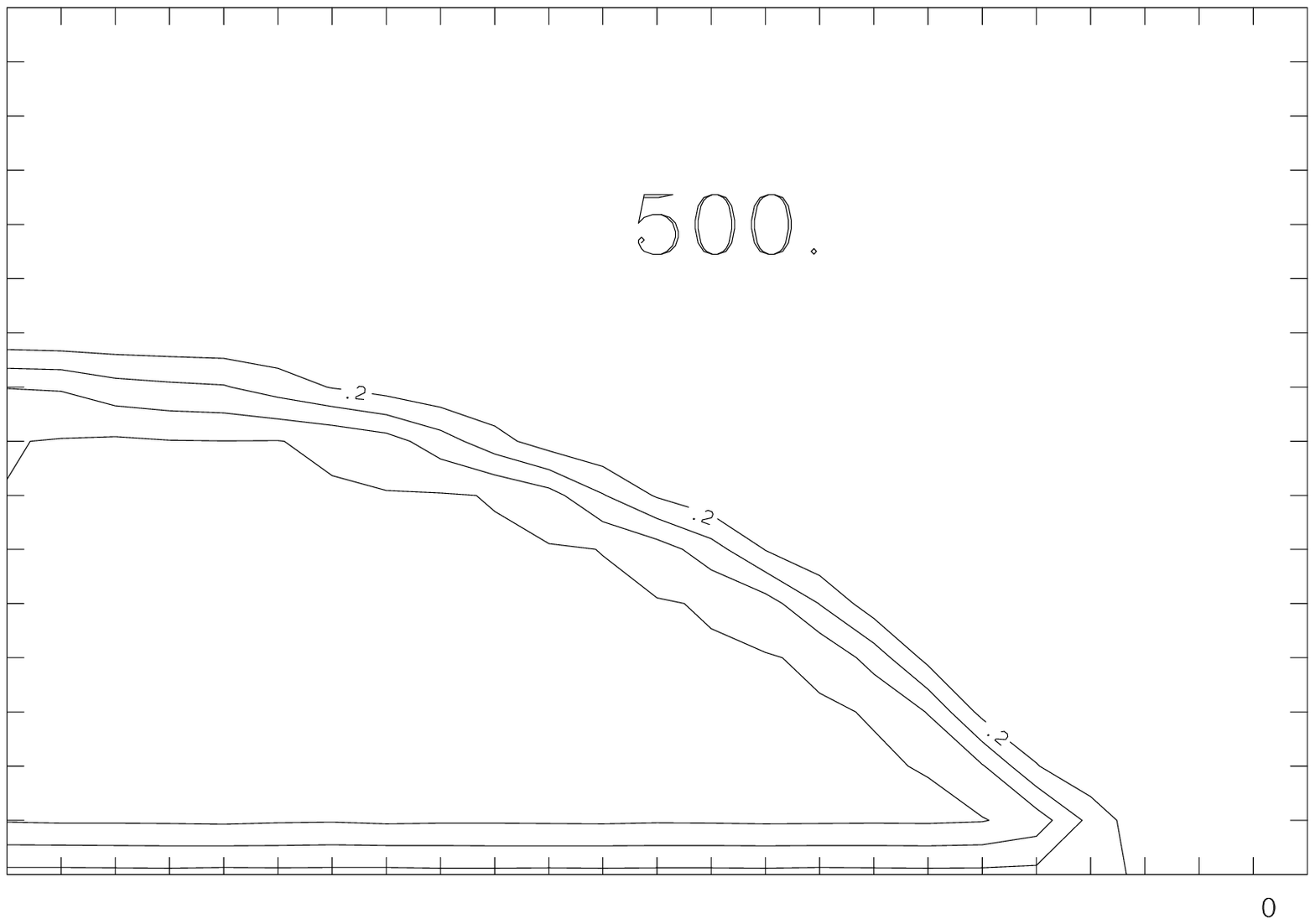}
  \caption{Density contours for a drop impacting a completely wetting
  surface at impact velocity 1.0 $\sigma/\tau$.  The format is identical to
  Fig.~\ref{fig2}.}
  \label{fig3}
  \end{center}
\end{figure}

At velocity 1.5 $\sigma/\tau$ the drop again bounces, but at intermediate
times develops a remarkable transient toroidal shape before surface tension
acts to restore it to a sphere: see the middle rows of
Figs.~\ref{fig1} and \ref{fig2}.  The density contours are similar to the
lower-velocity case up to time 100$\tau$ but then show the drop continuing 
to spread and a hole opening at the drop
center.  In comparison to the lower-velocity case,
after impact here the spreading lamella moves outwards so rapidly as to 
evacuate the central region about the impact point.  Aside from its
transience, 
the toroidal state is somewhat special since it only appears in a narrow
velocity range for tetramer drops and was not observed in simulations 
in the dimer systems.  An intuitive explanation is that a tetramer 
liquid is more coherent and less extendible than monomer or dimer liquids 
because the longer chains tend to intermingle.  When stretched outwards by
inertia, the drop resists indefinite thinning and disintegration by
rupturing to form a dense ring.

At still higher impact velocity, 2.0 $\sigma/\tau$, the drop again initially
evolves into a hemisphere and then again into a lamella, which is distinctly 
thinner than those at lower velocities. The lamella spreads but its
thickness and density decrease with time, and instead of contracting it
develops a splash and
eventually evaporates as the drop disintegrates  -- see the bottom rows of
Figs.~\ref{fig1} and \ref{fig2}.  The mean surface falls apart into isolated
liquid regions, meaning that lower-density fluid fills the simulation region.
A liquid jet is in fact emitted from the edge of the lamella (a prompt splash)
around time 30$\tau$, but this is a low-density phenomenon not evident
in these plots and will be illustrated below.  At even higher
velocity, 3.0 $\sigma/\tau$, the surface and density plots are similar (on a
faster time scale) but the drop seems to disintegrate without a distinct 
splash phase.  To elucidate splashing, and to compare the impact behavior
of different liquids, in the following subsection we employ an alternate 
display of the drops with molecular resolution.

\begin{figure}[t]
  \begin{center}
  \includegraphics[width=0.4\linewidth]{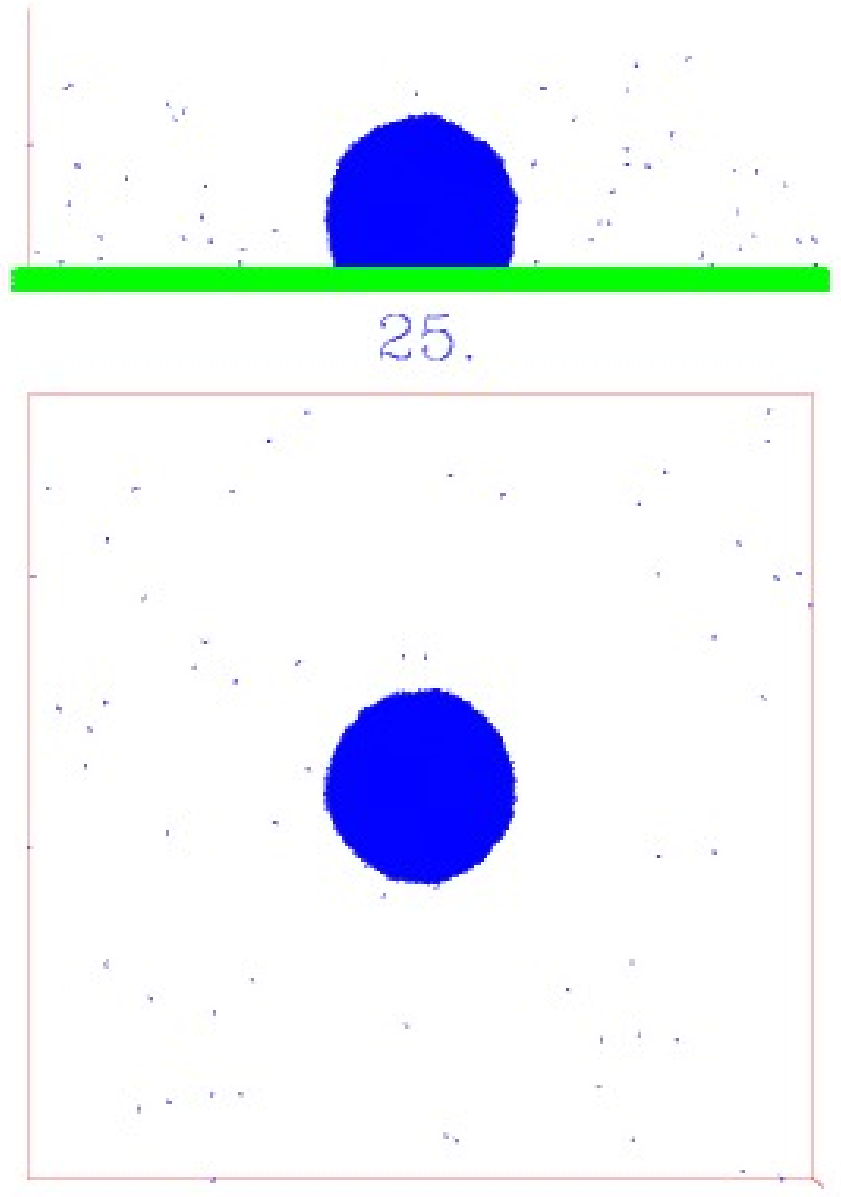}\hfill
  \includegraphics[width=0.4\linewidth]{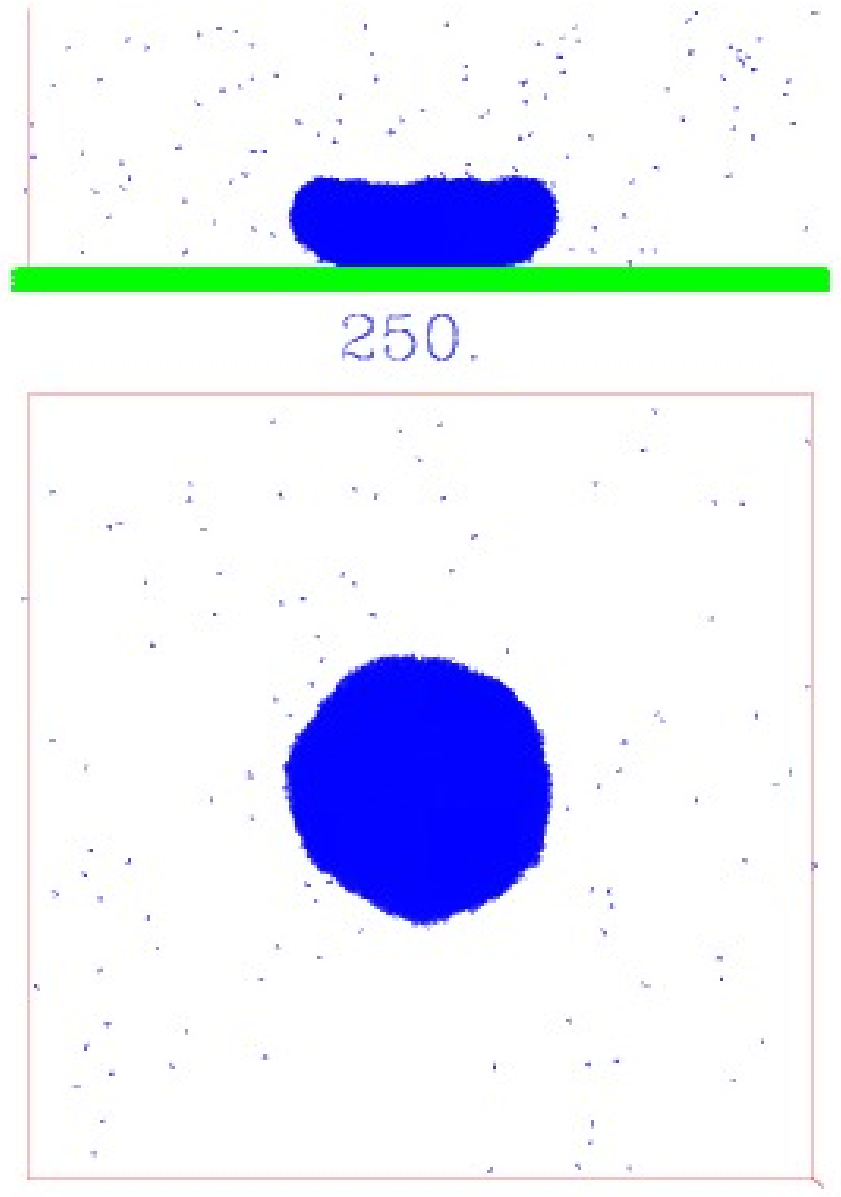}
  \caption{Simultaneous top and side molecular views of an 
  impacting tetramer 
  drop at velocity 1.0 $\sigma/\tau$ at the time indoicated. 
  Each molecule is displayed as the three 
  (blue) bonds joining the atoms and the horizontal (green) band indicates
  the positions of solid atoms.  Left: slightly after impact, Right: during
  recoil.}  
  \label{fig4}
  \end{center}
\end{figure}
  
In the bouncing cases, note that the thickness of the interfacial   
region -- the width of the set of parallel curves enclosing the drop -- does
not vary substantially during the bounce.  This indicates that the interface
does not broaden and that there is little density variation in the interior 
of the
drop, or in other words that the drop is approximately incompressible in this 
case.  The interfacial thickness here is slightly larger than that of a drop 
at rest, because the drop is in motion during the averaging time interval. 
The density fluctuates weakly during the
bounce, but the peak value (roughly 10\% above the value at rest) occurs 
only briefly and locally, just after contact in the hemispherical state.
In the splash regime the situation is quite different. There is no evidence 
for a constant density
interior during and after impact, but rather the density varies throughout
the drop's interior and furthermore decreases systematically during the 
splash and disintegration as more and more vapor appears. 

\begin{figure}[t]
  \begin{center}
  \includegraphics[width=0.4\linewidth]{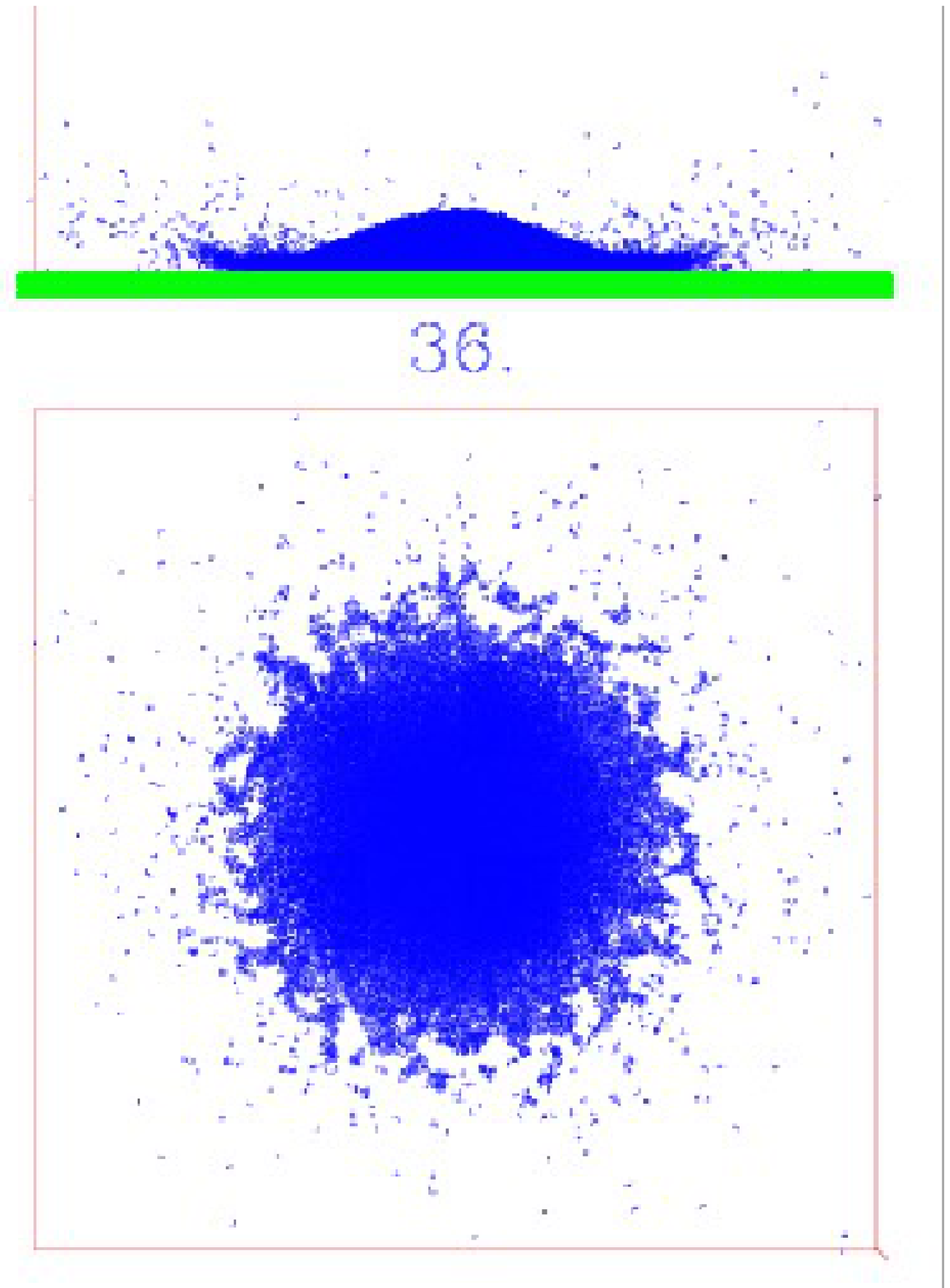}\hfill
  \includegraphics[width=0.4\linewidth]{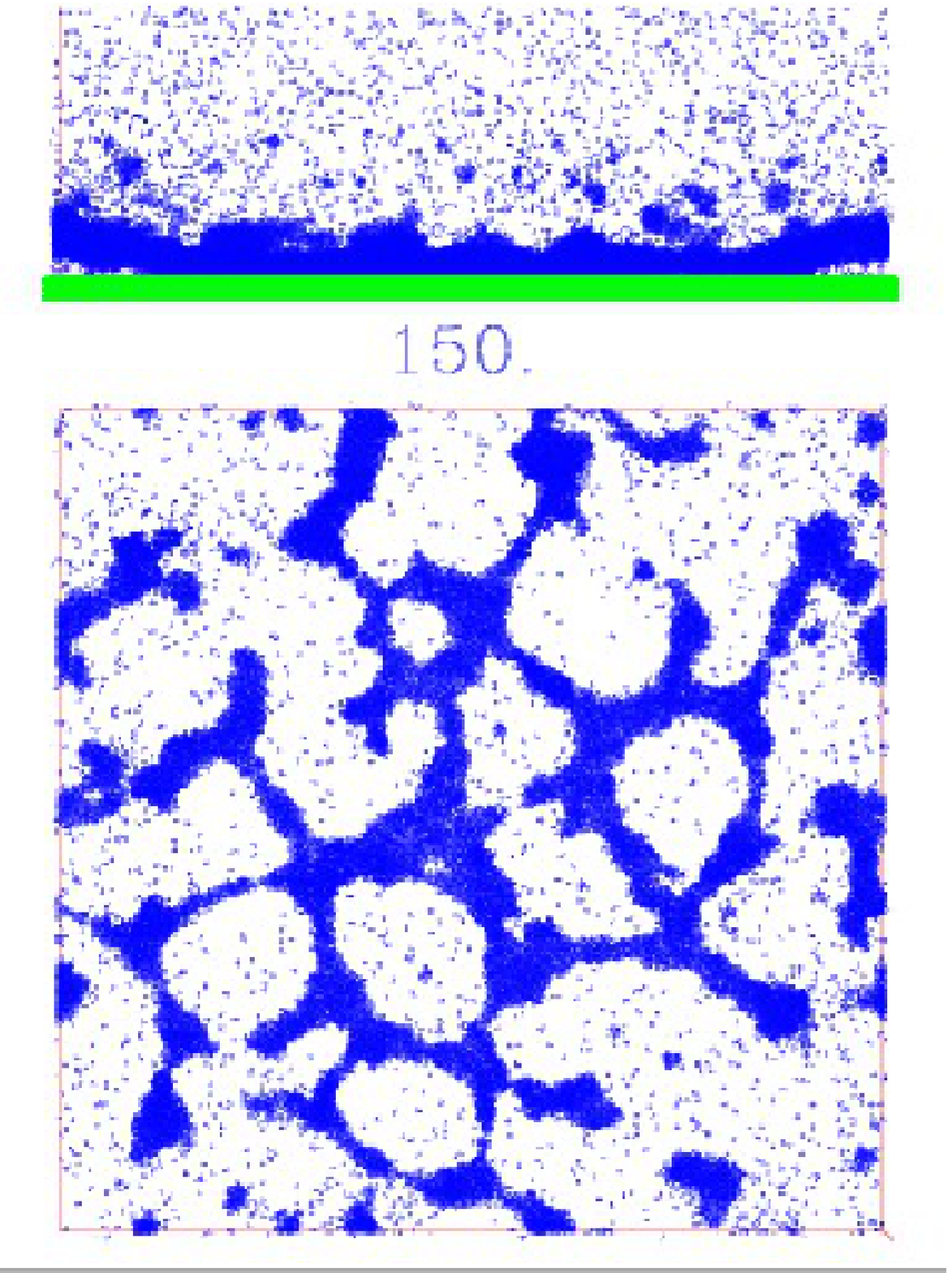}
  \caption{Simultaneous top and side molecular views of 
  an impacting tetramer
  drop at velocity 2.0 $\sigma/\tau$. Each molecule is displayed as the three
  bonds joining the atoms.  Left: during the splash phase, Right: at the end
  of the simulation.}
  \label{fig5}
  \end{center}
\end{figure}

The difference in drop behavior between wetting and non-wetting surfaces is 
indicated in Fig.~\ref{fig3}.  At time 100$\tau$, the spreading lamella on a
{\em completely wetting} surface has ceased its rapid outward motion, and its
tip has a downward curvature and a dynamic contact angle around 90$^\circ$.
At later times, the drop continues to spread due to capillary forces, but
this is a much slower process than inertially-driven spreading, and even after 
500$\tau$ little further outward motion has occurred.  In contrast,
on the non-wetting surface in Fig.~\ref{fig2}, in the top row at
time 100$\tau$ the tip of the lamella is detached from the surface with an
upwards curvature and a dynamic contact angle closer to 180$^\circ$.

\begin{figure}[t]
  \begin{center}
  \includegraphics[width=0.4\linewidth]{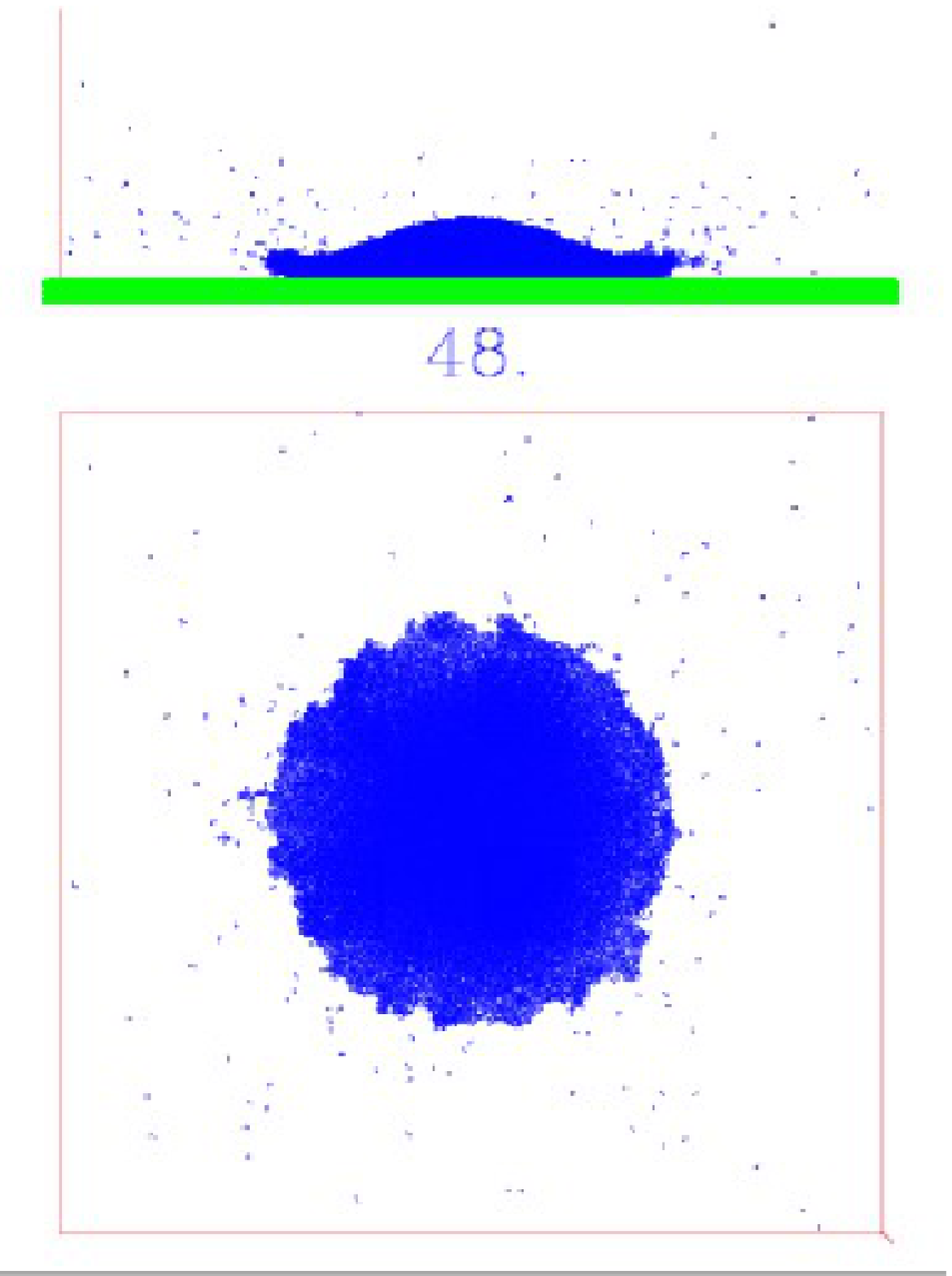}\hfill
  \includegraphics[width=0.4\linewidth]{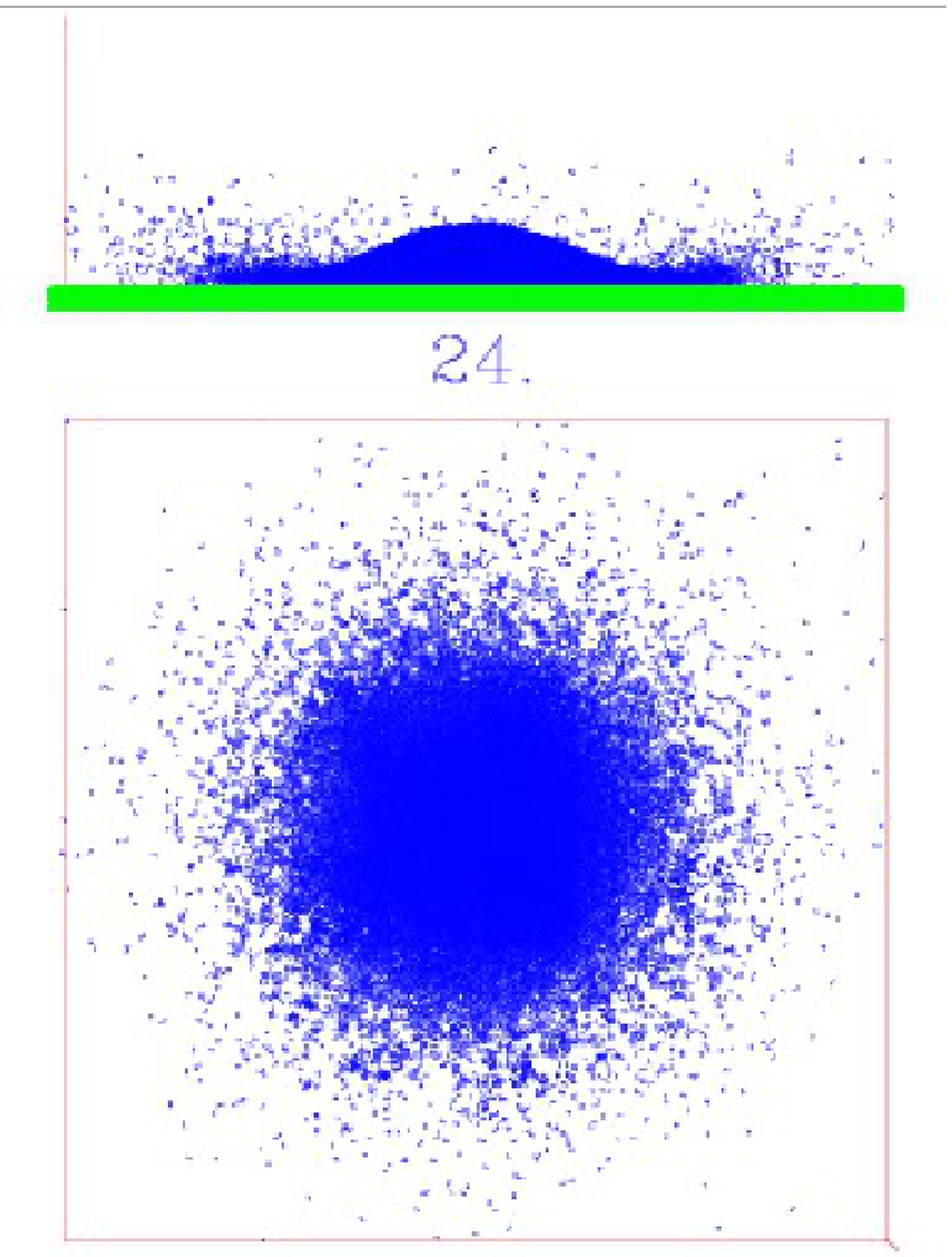}
  \caption{Molecular views for comparison with Fig.~\ref{fig5} 
  at 36 $\tau$.  Left: tetramer drop at velocity 
  1.5 $\sigma/\tau$ at time 48$\tau$, Right: tetramer drop at velocity 
  3.0 $\sigma/\tau$ at time 24$\tau$.  }
  \label{fig6}
  \end{center}
\end{figure}

\subsection{Molecular detail}

The distinction between the slow and fast impact cases and the effects of
external vapor are most evident when   
individual molecular positions are displayed. 
Figure~\ref{fig4} shows a non-volatile bouncing drop in simultaneous top and 
side views, at early
and late times after impact.  Some individual vapor molecules are
visible, but those inside the liquid are dense enough to uniformly fill the 
interior.  The impact has slightly increased the amount of
vapor but the liquid is quite coherent and the liquid-vapor interface and 
contact angle 
are well-defined.  Hence, the mean interface and density plots
above accurately capture the shape evolution of the drop.
At twice the impact velocity, however, we see in Fig~\ref{fig5} that as the 
liquid spreads on the surface, a significant number of 
molecules are emitted into the vapor from the drop rim region.
This is an example of a prompt splash:
at time 36$\tau$ there is a crown-like rim rising above the surface and 
droplets breaking off at roughly regular angular spacing around the edge. 
Subsequently, the crown collapses as the drop evolves into a single 
pancake-like lamella which continues to spread but becomes inhomogeneous
and develops holes, meanwhile steadily 
emitting vapor from its entire surface.
Eventually,  the liquid is reduced to a fragmentary pattern
resembling spinodal decomposition, while enough vapor is 
produced to almost uniformly fill the simulation box.
If the surface is completely wetting rather than repulsive, the
corresponding molecular views at this impact velocity are similar through
time 40$\tau$ or so but afterwards, while continuing to emit vapor, the 
lamella persists throughout the simulation as a well-defined thin disc or
puddle which does not disintegrate. 

\begin{figure}[t]
  \begin{center}
  \includegraphics[width=0.4\linewidth]{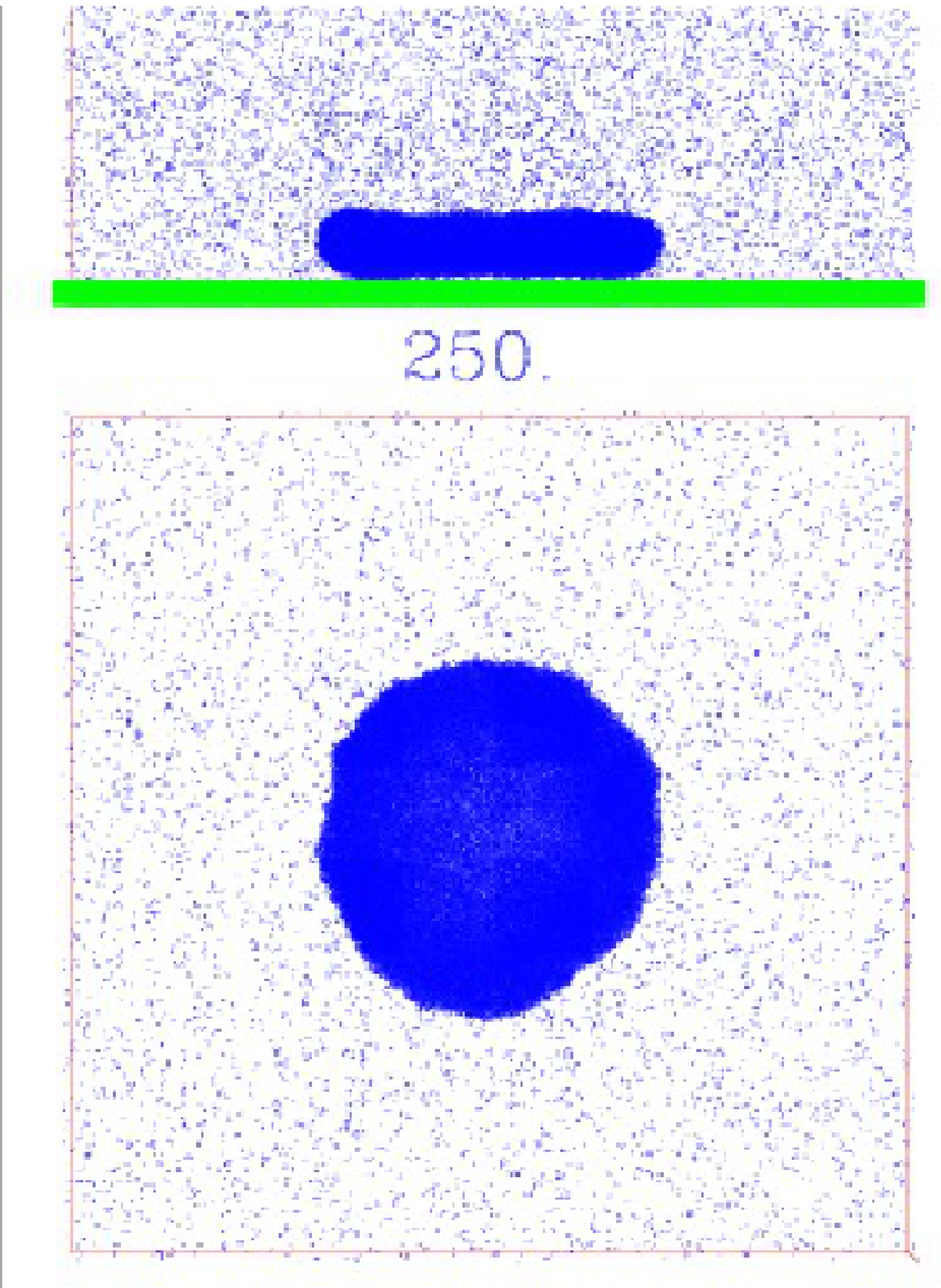}\hfill
  \includegraphics[width=0.4\linewidth]{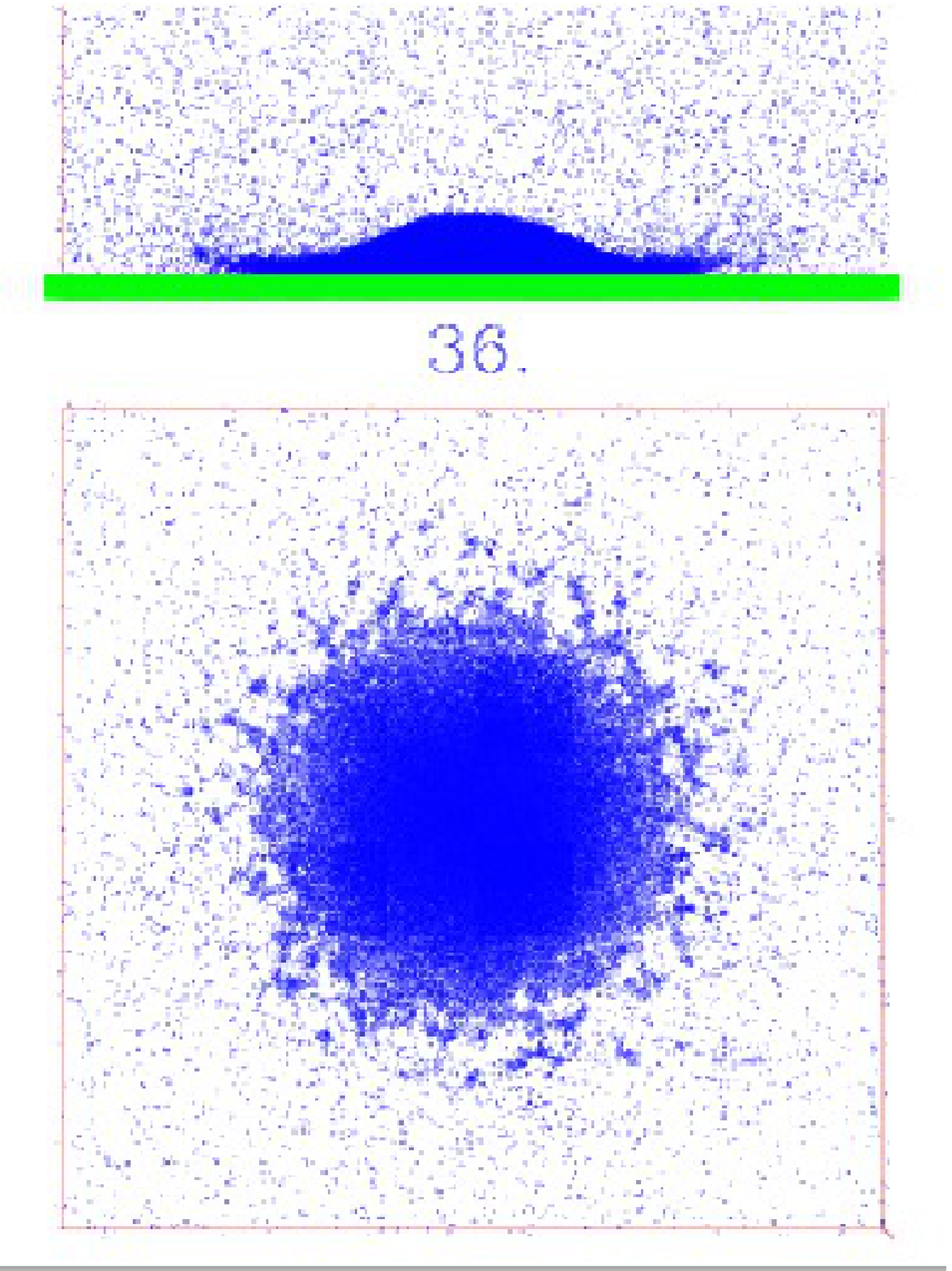}
  \caption{Molecular views of an impacting dimer drop at 
  velocity 1.0 (left) and 2.0 $\sigma/\tau$ (right), at the same times as 
  for the tetramer drops in Figs.\ref{fig4},\ref{fig5}.}
  \label{fig7}
  \end{center}
\end{figure}

The splashing state should be distinguished from drop disintegration, which
occurs at still higher impact velocities.  In Fig.~\ref{fig6} we show
comparison 
snapshots of non-volatile impacts at both lower and higher impact velocities
at times corresponding to the same displacement $u_0t$ as Fig.~\ref{fig5} at 
time 36 $\tau$:  $u_0$=1.5 $\sigma/\tau$ at $t$=48 $\tau$ and
$u_0$=3 $\sigma/\tau$ at $t$=24 $\tau$. The slower case -- the toroidal
bounce -- has a slightly irregular rim but no significant vapor emission as
the lamella spreads.  In the higher-velocity case, 
vapor is emitted from the entire lamella while the
edge of the rim has a radially-decreasing density with little angular 
variation.  
The final state in this case (not shown) has a few irregular droplets on the
surface and otherwise a dense fluid filling the box. 
 
A key result of this paper is that this behavior is generic:  the choice of
molecular liquid and the presence 
or absence of surrounding vapor does not qualitatively effect the dynamics 
of nano-sized drop impacts.  This conclusion is based on repeating these 
simulations with dimer liquids, with vapor present initially or not, and
mixed tetramer-liquid/dimer-vapor drops with
the same initial radius and impact velocities as those above. In all cases
she mean-surface plots and density fields are very similar to those of 
tetramers and the only distinction is in the amount of vapor present after 
impact.  For example, in Fig.~\ref{fig7} we display molecular snapshots of the
dimer liquid at impact velocities 1.0 and 2.0 $\sigma/\tau$ at two of
the same times
as the tetramer simulations shown in Figs.~\ref{fig4}b and \ref{fig5}a,
respectively.  In the first case, the drop is bouncing and has a
well-defined and relatively smooth interface, with rather more vapor outside
the drop.  In the second frame the dimer drop is disintegrating in essentially 
the same way as the tetramer, and the vapor molecules are simply floating 
in the background.

\begin{figure}[t]
  \begin{center}
  \includegraphics[width=0.5\linewidth]{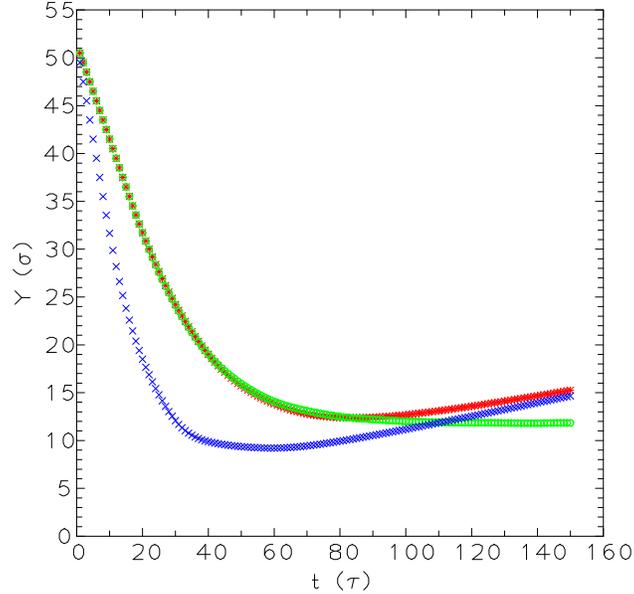}
  \caption{Variation of the height of the center of mass of 
  the fluid (liquid plus vapor) as a function of time.  
  Initial velocity 1.0 (\red{$*$}), and 2.0 $\sigma/\tau$ 
  (\blue{$\times$}) on a non-wetting surface and velocity 
  1.0 $\sigma/\tau$ (\green{$\circ$}) on a wetting surface.}
  \label{yoft}
  \end{center}
\end{figure}

Of course, variation of the liquid properties does change the Reynolds and
Weber numbers for impact at a given velocity, and strictly speaking we
should compare the different liquids by matching these quantities. In the
examples above, the operating conditions were well inside the bounce and
splash regimes, respectively, and the distinction was not crucial, but a
difference does occur near the transition values.  At velocity
1.5 $\sigma/\tau$ for example, the tetramer drop bounces (via the toroidal
intermediate state shown in Fig.~\ref{fig1}) but a dimer drop at the same
velocity is found to simply splash. The dimer has higher Reynolds and Weber 
numbers and this behavior is still consistent with a critical Weber number
O(100) dividing the bouncing and splashing regimes. The experimental
transition Weber number is rather higher than this, O(500), which is one
reason we speak of disintegration rather than splash -- see below.

The insensitivity to vapor is in contrast to the behavior of larger, 
millimeter-sized drops, where vapor appears to form a lubrication layer which
causes a splash at sufficiently high velocity.  The key difference is that
although the vapor density in theses simulations is similar to room
conditions, the amount of vapor lying beneath a falling 12 nm 
drop is very small - tens of molecules - and insufficient to show any
significant hydrodynamic behavior.  When a drop impacts the surface, these
few molecules are either adsorbed into the drop or pushed aside but do not
effect its behavior.  In further support of this conclusion, we have
carried out two types of modified simulation.  First, we
simulated the impact of dimer drops in which either a fraction or all of the
vapor molecules are deleted at the instant the drop is set into motion:
there is no significant change in the splash.  Second, in a complementary
simulation, we surrounded a tetramer drop with dimer vapor before directing
it downward.  While the shape of the drop is slightly altered, Again there 
was no significant change in comparison to the original system.

\begin{figure}[t]
  \begin{center}
  \includegraphics[width=0.3\linewidth]{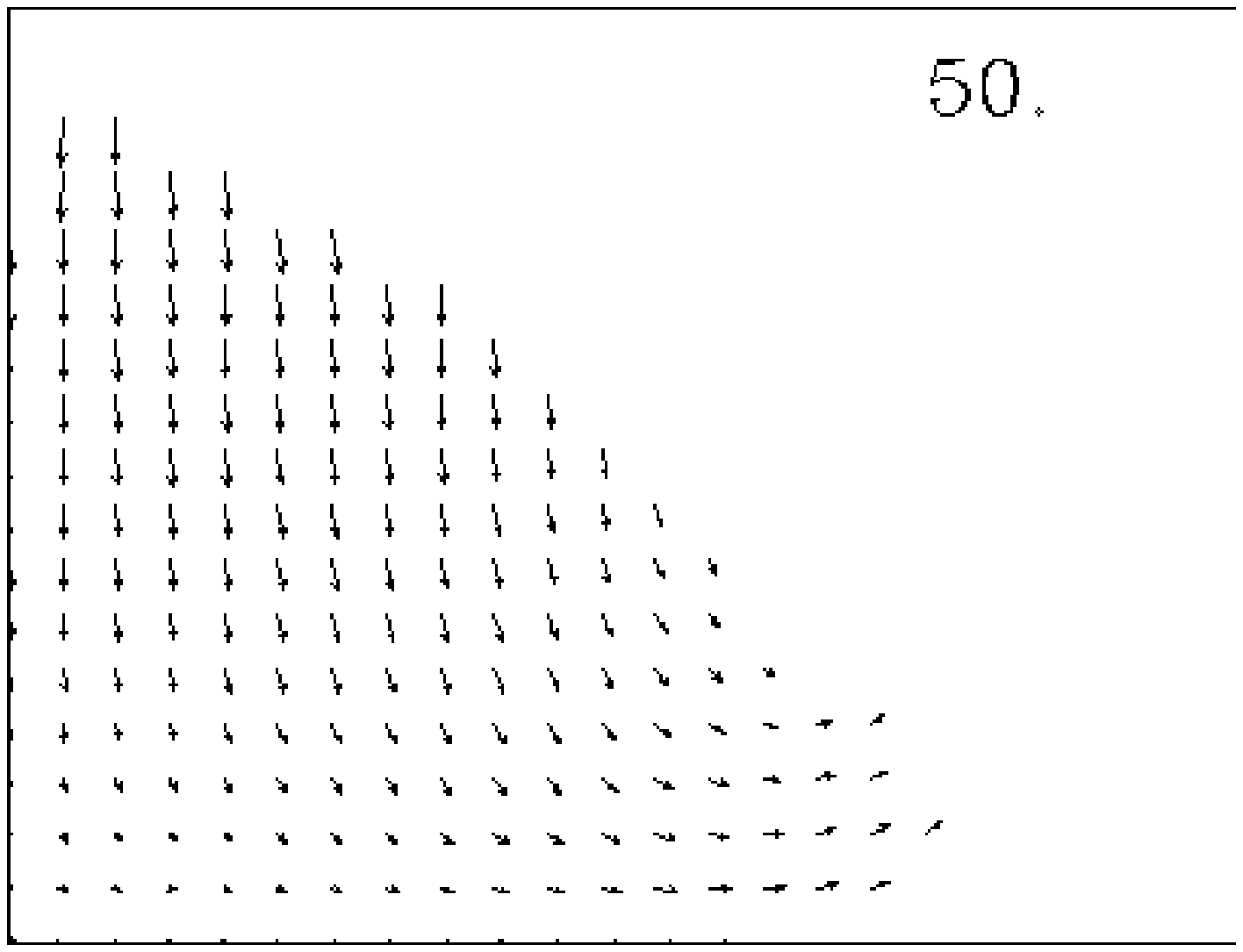}
  \includegraphics[width=0.3\linewidth]{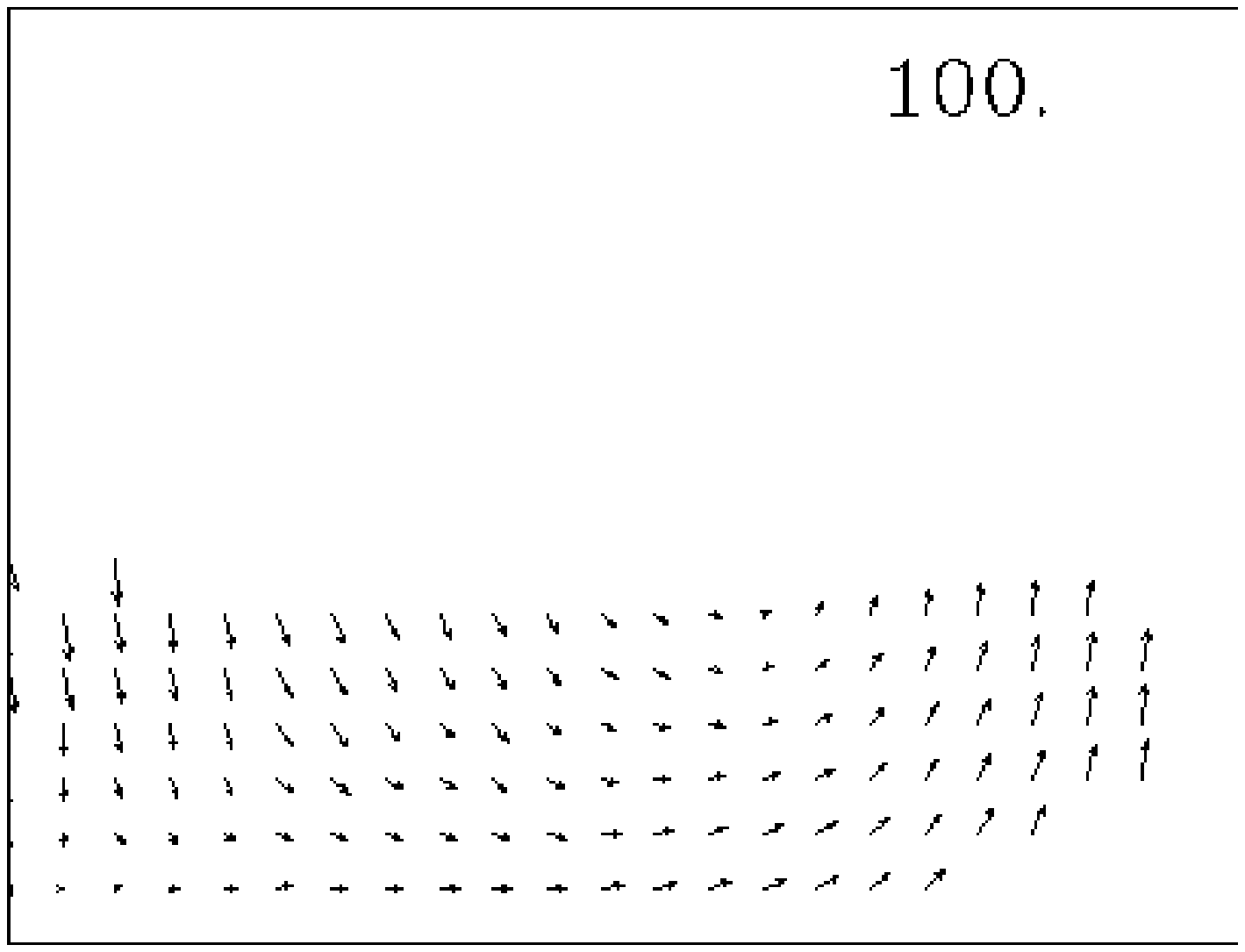}
  \includegraphics[width=0.3\linewidth]{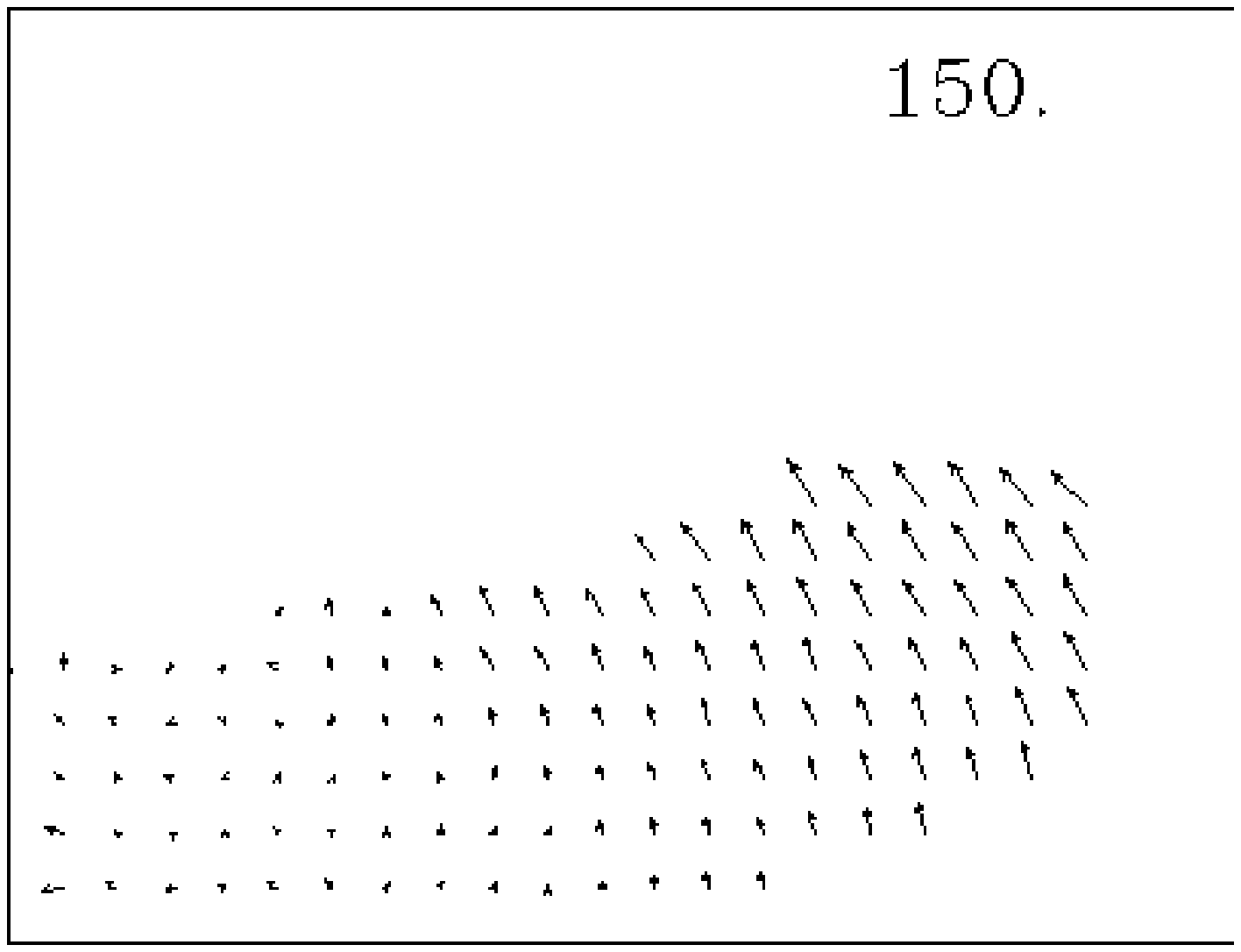}
  \caption{Cylindrically averaged velocity field for a non-volatile drop
  impacting a surface at $u_0$=1.0 $\sigma/\tau$.  Each arrow represents the
  velocity averaged over a 3$\sigma\times3\sigma\times360^\circ$ spatial
  bin and a 10$\tau$ time interval.  The maximum velocity (longest arrow)
  in each frame is 0.94, 0.24 and 0.13 $\sigma/\tau$, respectively.}
  \label{fig8}
  \end{center}
\end{figure}

\subsection{Flow fields}

The overall motion of the drop relative to the surface 
is not completely clear in the preceding
figures, but may be shown by plotting the height of the center of
the drop (more precisely, $Y(t)$, the $y$-coordinate of the center of mass 
of the fluid atoms) as a function of time.  The result for tetramer drops is
shown in Fig.~\ref{yoft}:  a V-shaped curve when the drop
bounces on a non-wetting surface and an L-shaped curve when the drop sticks to 
a wetting surface. The curve is always asymmetric because energy is adsorbed
by the (thermostated) wall atoms during impact and spreading.
A drop which splashes and then disintegrates also shows a
nearly V-shaped variation, corresponding to the liquid drop moving downwards
with the initial velocity followed by an smooth rise of the center of mass
when its fragments rebound upwards off the surface.  Once again, the
corresponding curves for dimer drops are quite similar. 

In addition to the density field shown above, we have attempted to measure 
two-dimensional fluid velocity and stress fields inside the drop.  As an
example, the cylindrically-averaged velocity field 
for a non-volatile drop impacting at 1.0 $\sigma/\tau$ is shown in 
Fig.~\ref{fig8}.  At 50 $\tau$ there is a roughly hyperbolic flow in which
the original bottom of the drop is forced outwards while the upper half 
is still moving downwards.  The radially outwards flow produces the
spreading lamella, but at 100 $\tau$ the radial motion at
the edge of the rim has ceased and the rim is bending upwards, while the 
top of the drop continues to fall, and at 150 $\tau$ the velocities are
very small in the center of the drop while a faster contraction to a sphere is
underway at the rim.  The same sequence is observed for all bouncing drop 
cases, whereas when the surface is wetting the velocities simply decay to 
small random 
fluctuations when the initial rapid-spreading stage ends. In splashes or
disintegrations, the velocity field is simply radially outwards at later 
times before decaying away.  In these figures the vapor region is not shown 
because it contains too few molecules for
averaging to produce a robust signal above the statistical fluctuations.

The pressure contours are similar in overall shape to the density field,
meaning there is a pressure where there is a (dense) liquid, and beyond this
the pressure is high at the wall due to impact. It is difficult to go beyond
these qualitative statements here because of numerical imprecision. 
The shear stress is simply too noisy to resolve a signal above the
fluctuations here.  Generally, the
stress tensor fluctuates strongly in molecular dynamics simulations,
because it directly involves the inter-atomic force, which is a rapidly
varying function of position.  The use of long intervals of slowly-varying
behavior for time-averaging or 
large homogeneous regions for spatial averaging can produce a
robust signal, but these are absent in this problem.  We have attempted to
average the results over a modest statistical ensemble (of five realizations)
to no avail.  Presumably a much larger drop or much slower
velocity would be required for precise stress measurements. 

\section{Discussion}
\label{sec:disc}
These results indicate that
splashing nanodrops differ from their larger-sized counterparts in (at
least) two significant ways. First, the vapor surrounding a drop does not
control the occurrence of splashes, basically because even though its density 
is comparable to laboratory values there are too few vapor molecules 
beneath a nano-sized drop to have any significant effect.  In contrast, 
a macroscopic
drop falling through an appreciable distance can trap a continuum air film
beneath itself, which will build into a pressurized lubrication layer as 
the drop nears
the surface, which in turn can exert enough shear stress on the bottom of 
the drop to deform it. Here we have only tens of molecules and no real film.
The second difference is in the nature of the splash itself - at high
velocities the edge of a nanodrop deformed by rapid impact appears to spew 
out molecules and disintegrate, and eventually the entire drop seems to
fall apart.  In contrast, macroscopic drops have enough
molecules to retain their identity as dense liquids even after substantial
deformation and stretching of the liquid-vapor interface, and even in such 
dramatic processes as crown formation the rim of the drop is still just a
liquid sheet with bulbous  protuberance at its edge.  

One reason for the different structure in the final state of nanodrop splashes
is volume -- there are too few molecules available to populate all 
of the fine liquid details of a continuum splashing drop.  A second reason is 
related to the high molecular speeds $u_0 \sim 1.0\, \sigma/\tau \sim 100$m/s
necessary for a nano-sized drop to reach Reynolds and Weber numbers 
O(100). In confined flows 
such velocities are not problematic, providing that the results are
understood in terms of the appropriate dimensionless velocity {\em gradient}. 
if the Deborah or Weissenberg number is small one is simulating a Newtonian 
fluid and otherwise non-Newtonian behavior is expected.  When a
liquid-vapor interface is strongly deformed, however, there is the issue of
coherence -- holding the liquid together. 
Some insight into the lack of liquid coherence in the high-velocity impacts
here comes from simple energetics:  when the drop is at rest the potential 
energy per atom due to the LJ interaction is measured here to be 
-1.35$\epsilon$ and -1.48$\epsilon$, respectively, in the dimer and tetramer 
cases.  Since the kinetic energy per atom is 
${\textstyle{3\over 2}} k_BT = 1.2\epsilon$, 
the average energy per atom is only slightly negative, and since the
interatomic potential vanishes outside a cutoff, in a sense the liquid
is loosely bound.  The kinetic energy is significantly increased by the 
initial velocity, nominally by 4.5 $\epsilon$ in the fastest case above, 
and on impact the outward motion produces large velocity gradients and
therefore large relative kinetic energies of interacting pairs of molecules,
which may then become unbound.  Laboratory drops do not have so high an
impact velocity and this effect is absent.  This argument is not 
precise enough to accurately predict the transition velocity in nanodrop 
splashing, it is consistent with the observed  transition velocity of about 
1.5 $\sigma/\tau$ and it does help explain the nature of the final state.

An alternative way of thinking about drop disintegration is in terms of
temperature.  In the low impact velocity case, the drop is initialized at
$T$=0.8 $\epsilon/k_B$, which does not change when the drop is translated as a
body (temperature is proportional to the mean-square velocity {\em fluctuation} 
about the mean), but the impact produces non-uniform velocity and
temperature fields. At lower impact velocities the local temperature at the
edge of the drop varies from 0.8 to 1.0, but in the higher velocity impact
values above 2 occur.  The point is that the latter values are above the
critical temperature for the liquids simulated here, and since there is no   
distinction between liquid and vapor phases the drop is evaporating.  
Admittedly,
the phase coexistence curve is directly meaningful only for homogeneous
systems in equilibrium, but it does provide some qualitative insight.
For a pure monatomic LJ fluid the critical temperature is known to be 
1.1 $\epsilon/k_B$ \cite{watanabe}, 
but the molecular binding is expected to change the value.
While we have not attempted to determine the full phase diagram, we have
estimated the maximum temperature for phase coexistence at the densities 
of the present simulations by placing the equilibrated drop plus 
vapor molecular configurations in a periodic box of the same dimensions
used in the impact calculations and observing the behavior at various
temperatures.  The result is that dimer drops evaporate at $T\simeq 1.1$ and
tetramers at $T\simeq 1.4\epsilon/k_B$, and since we observe that
these temperatures are exceeded on fast
impact, we could qualitatively describe the drop disintegration seen here as 
rapid evaporation.

\begin{acknowledgments}
We thank M. Brenner, S. Nagel and T. Witten for helpful discussions. 
\end{acknowledgments}

\newpage
\appendix
\section{Sound Speed}

\begin{figure}
  \begin{center}
  \includegraphics[scale=0.45]{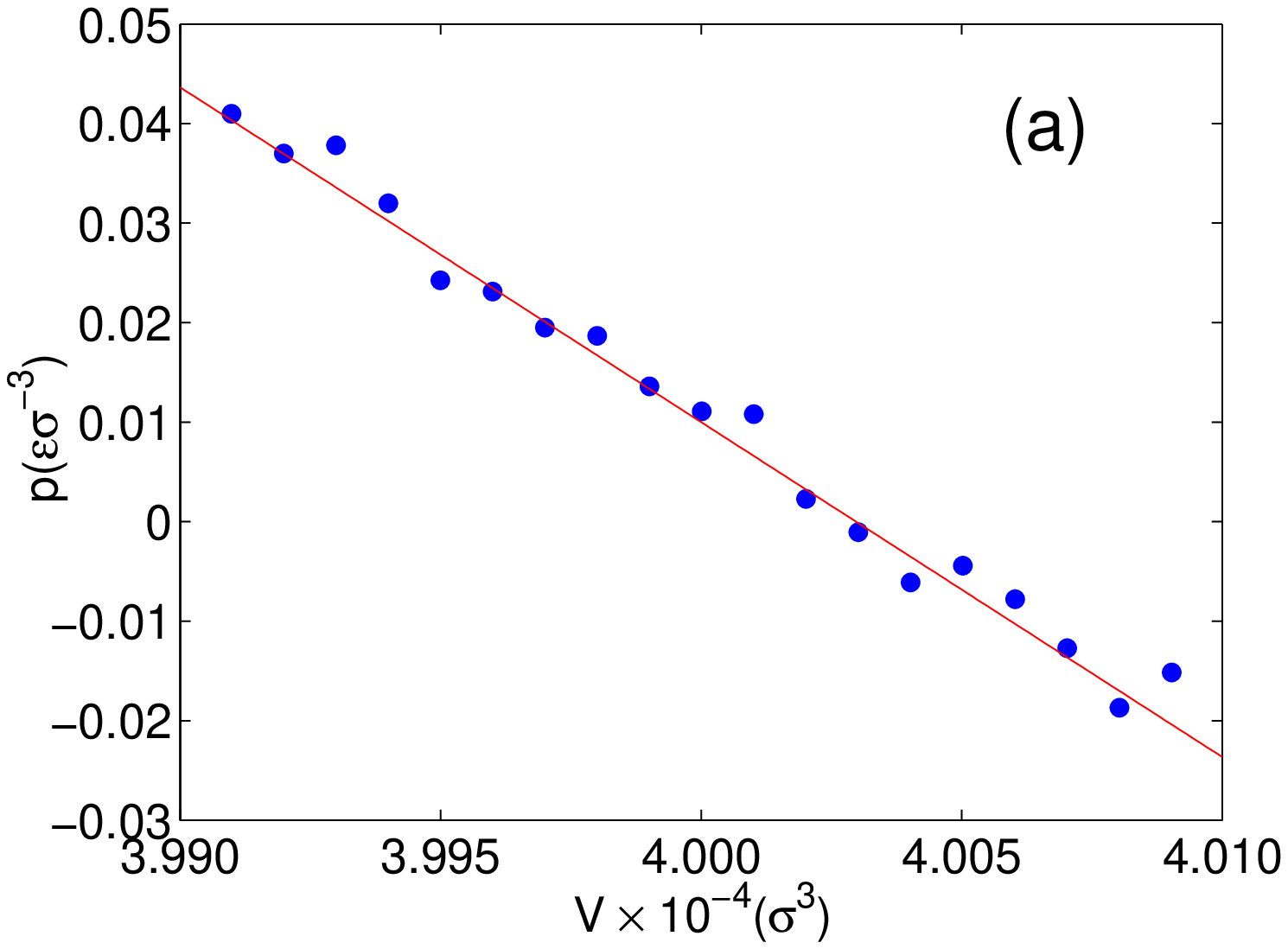}
  \includegraphics[scale=0.45]{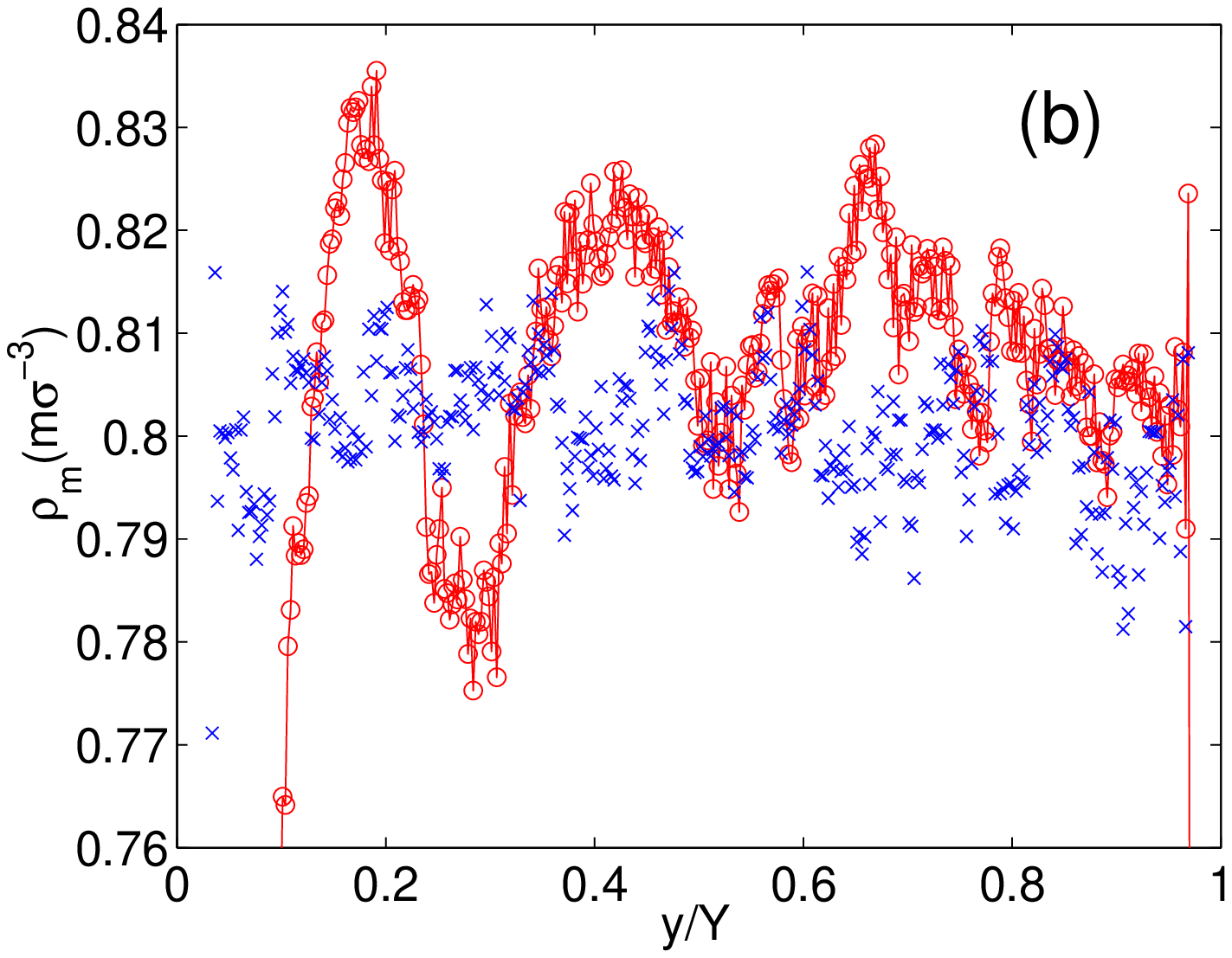}
  \caption{(a) $p$-$V$ diagram for dimers;  points (blue) are
  simulation results and the line (red) is a fit.  (b) Density profile
  produced by an oscillating wall: {\red{$\circ$}} oscillating wall,
  {\blue{$\times$}} stationary wall.}
  \label{pV-ry}
  \end{center}
\end{figure}

The sound speed is given generally by
$$
u_s^2=\left(\frac{\partial p}{\partial \rho}\right)_S
=-\frac{V}{\rho} \left(\frac{\partial p}{\partial V}\right)_S,
$$
where $p$ and $\rho$ are the pressure and mass density of the 
medium, respectively, where $M$ is the atomic mass. The second form is 
useful in MD calculations, where it can be
implemented by a sequence of NVT simulations in which the volume $V$ is
slowly varied while the constant entropy ($S$) constraint is satisfied by
isolating the system thermally so that there is no heat flux.  We first
equlibrate a system of 16,000 dimer molecules in a periodic cube of side
34.2$\sigma$ at temperature 0.8 $\epsilon/k_B$ and density either
slightly above or slightly below the target value 0.8 $m\sigma^{-3}$.
The volume is then either decreased or increased by 0.5\% over a 400 $\tau$
time interval, approximately an adiabatic variation in volume, so as to 
bracket the target density.  The resulting pressure variation is shown in
Fig.~\ref{pV-ry}a, from which the slope of the linear least-squares fit gives 
$u_s=(4.1\pm0.1)\sigma/\tau$.

As a check on the result we use a second method:  direct simulation of 
a sound wave.  In this case we place a dimer liquid at the desired density in
a rectangular simulation box of dimensions
$(X,Y,Z)=(20.5,215.5,20.5)\sigma$, bounded by two solid walls in the
(long) $y$-direction.  After equilibration, one wall is rigidly
oscillated in $y$ at frequency $\omega = 0.6 \tau^{-1}$
and amplitude $\sigma$. The opposite wall is fixed in place and 
thermostatted to adsorb the energy flux resulting from the oscillation.  
After a transient period, we observe a traveling wave of
spatial variation in density, depicted at one time in Fig.~\ref{pV-ry}b.
For reference, the density variation due to the oscillation is compared with 
that resulting from equilibrium fluctuations in the same fluid in a fixed
volume.
A clear signal is present near the moving wall, although the wave decays
near the fixed wall due to energy adsorption there.  From the spatial
variation of density a wavelength can be extracted and, given the frequency, 
we find $u_s=(4.8\pm0.3)\sigma/\tau$.  The two methods are in qualitative
agreement, but we consider the first to be more reliable.  Some deficiencies
of the second method are (1) the temperature of the liquid is not controlled, 
(2) we have not taken account of any reflected wave, and (3) a rather high 
frequency is needed in order to have a wavelength well below the size of the
simulation box.

\begin{table}[h]
\begin{center}
\begin{tabular}{|l||c|c|c|}
\hline
liquid  &  $\rho$ $(m\sigma^{-3})$& Method I  & Method II    \\ \hline\hline
monomer &  0.8			 &$5.0\pm0.1$  & $5.7\pm0.6$ \\ \hline
dimer   &  0.8			 &$4.1\pm0.1$  & $4.8\pm0.3$ \\ \hline
tetramer&  0.86			 &$4.8\pm0.2$  & $5.5\pm0.4$ \\ \hline
\end{tabular}
\end{center}
\caption{Measurements of the speed of sound at $T=0.8\epsilon/k_B$.
Method I is based on the definition and Method II  
involves the direct simulation of sound waves.}
\label{table3}
\end{table}

The measurements were repeated for a monomer Lennard-Jones fluid as well as
the tetramer fluid studied in the paper, and the results are summarized in
Table III.  The point of the monomer measurement is that this simulated 
material is a reasonable approximation to liquid Argon, whose sound speed has 
been measured experimentally to be 853m/s at 44.4MHz and 85K and 
atmospheric pressure \cite{sound}, close to the simulation result for Method
I.  The fact that the sound speed in dimer liquids is lower than in monomer
liquids is consistent in trend with ideal gas theory where 
$u_s=(\gamma k_B T/m)^{1/2}$:  the adiabatic index $\gamma$ is lower for 
dimers than monomers.

\end{document}